\documentclass{aa}  

\usepackage{natbib}
\bibpunct{(}{)}{;}{a}{}{,}

\usepackage{graphicx}

\usepackage[varg]{txfonts}
\usepackage{xcolor}
\definecolor{darkblue}{rgb}{0.0, 0.0, 0.5}
\usepackage[colorlinks=true, citecolor=darkblue, linkcolor=darkblue, urlcolor=black]{hyperref}

\begin{document} 

\title{Atmospheric evolution through outgassing and escape \\ on young molten rocky exoplanets}

\author{Emma Postolec\inst{1}\corrauth{e.n.postolec@rug.nl} 
\and Tim Lichtenberg\inst{1} \email{tim.lichtenberg@rug.nl}
\and Harrison Nicholls\inst{2,3} \email{harrison.nicholls@ast.cam.ac.uk}
\and Laurent Soucasse\inst{4,5} \email{laurent.soucasse@imec.be}
\and Floris van der Tak\inst{1,6}\email{f.f.s.van.der.tak@sron.nl}
  }

\institute{Kapteyn Astronomical Institute, University of Groningen, P.O. Box 800, 9700 AV Groningen, The Netherlands
\and
Institute of Astronomy, University of Cambridge, Cambridge, CB3 0HA, United Kingdom
\and
Atmospheric Oceanic and Planetary Physics, University of Oxford, Oxford, OX1 3PU, United Kingdom
\and 
Netherlands eScience Center, Science Park 402 (Matrix THREE) 1098 XH Amsterdam, The Netherlands
\and 
Interuniversity Micro Electronics Centre, Kapeldreef 75, 3001 Leuven, Belgium
\and 
Space Research Organisation Netherlands, Landleven 12, 9747AD Groningen, The Netherlands
}

\date{Received 5 February 2026 / Accepted 6 July 2026}

\abstract
{The earliest rocky planet atmospheres are shaped by the competition between initial volatile inventories and atmospheric escape driven by stellar irradiation. On young magma ocean planets, volatile exchange between atmosphere and interior through in- and outgassing competes with atmospheric escape, controlling volatile retention and atmospheric evolution.}
{We investigate how atmospheric escape and replenishment via outgassing during magma ocean crystallization shape rocky exoplanet atmospheres.}
{We extend a coupled interior-atmosphere model to simulate rocky planet evolution during the magma ocean era by incorporating an energy-limited atmospheric escape module. Comparing self-consistent radiative-convective and prescribed-convective atmospheres, we quantify how atmospheric energy transport affects escape. We explore a wide range of orbital separations, escape efficiencies, oxidation states, and initial volatile inventories to identify regimes where sustained magma-ocean outgassing or escape dominates. We estimate atmospheric mass loss and compositions for young rocky planets around Sun-like and M-dwarf stars over geologic timescales.}
{Atmospheric escape shortens magma ocean lifetimes by weakening greenhouse insulation and accelerating cooling, with radiative-convective atmospheres reducing solidification timescales compared to purely convective cases. Volatile dissolution into the magma ocean interacts with escape of atmospheric species to chemically fractionate the planetary volatile budget over time by retaining more soluble species. For Earth-mass planets, atmospheres survive if mass-loss rates remain moderate ($\dot{M}_{\rm EL}\approx 10^{10}\,\mathrm{g\,s^{-1}}$). When atmospheric escape is included, mantle redox state remains a key control on atmospheric composition: high oxygen fugacity ($f$O$_2$) yields heavier, \element{H_2O}- and \element{CO_2}-rich atmospheres, while low $f$O$_2$ produces light, \element{H_2}- or \element{CO}-dominated atmospheres, consistent with previous studies. Orbital separation, initial volatile inventory, and stellar type produce diverse evolutionary pathways, from bare rocky planets to magma oceans with thick atmospheres, spanning compositions from \element{H_2}- to \element{SO_2}-dominated atmospheres.}
{}


\maketitle
\nolinenumbers  

\section{Introduction}

Atmospheric compositions of gaseous and sub-Neptune exoplanets have been now characterized thanks to the spectroscopic capabilities of the \textit{James Webb Space Telescope} \citep[JWST--][]{Tsai_2023,Benneke_2024,Dyrek_2024,Piaulet-Ghorayeb_2024,Wallack_2024,wogan2024}. However, observing the atmospheres of smaller, rocky planets such as super-Earths and Earth analogs remains challenging due to their small sizes and weaker spectral signatures \citep{May-2023,Moran_2023,kreidberg_stevenson_2025}. Despite several attempts, spectral characterization of super-Earth secondary atmospheres are still at a tentative stage, like 55\,Cancri\,e \citep{Mercier_2022,Hu_2024,Patel-2024} or TOI-561\,b \citep{teske_2025}. Some exoplanets orbiting M dwarfs, such as GJ\,1132\,b \citep{Schaefer_2016, Diamond-Lowe-2018, Libby-Roberts-2022, May-2023, Xue-2024}, TRAPPIST\,1\,b \citep{Greene_2023, Ducrot_2025}, TRAPPIST\,1\,c \citep{Zieba_2023}, and GJ\,486\,b \citep{Moran_2023,mansfield2024thickatmosphereterrestrialexoplanet} remain ambiguous: both a bare-rock interpretation and models with thick atmospheres can explain current observations \citep{Coy2025ApJ,Hammond_2025}. This ambiguity underlines the challenge of obtaining conclusive observational evidence for secondary atmospheres around rocky exoplanets with current instrumentation. GJ\,367\,b \citep{Zhang2024ApJL}, TOI\,1685\,b \citep{luque2024darkbarerocktoi1685}, and LHS\,3844\,b \citep{Kreidberg_2019} are consistent with a bare rock scenario, suggesting their atmospheres may have been removed by strong stellar irradiation history over evolutionary timescales. However, the increasing evidence for secondary atmospheres on ultra-short period exoplanets \citep{Hu_2024,teske_2025} speaks against stellar irradiation as a simple explanation of atmospheric loss or retention. These contrasting cases highlight the uncertainty surrounding the long-term atmospheric retention of rocky planets over geological timescales \citep{kreidberg_stevenson_2025,Lichtenberg_shorttle_teske_kempton_2025}.

During their early evolution ($10^6$--$10^8$\,years), rocky planets host a global magma ocean containing volatile species such as \element{H_2O}, \element{CO_2}, and \element{SO_2} dissolved into the molten silicate layer due to their high solubilities at extreme temperatures ($\gtrsim1500$\,K) and pressures \citep[up to several GPa;][]{SOSSI_2023, Suer_2023}. As the planet cools over time, these volatiles are outgassed, enriching the primary atmosphere \citep{Gaillard_2022, 2023_lichtenberg_pp7}. However, when those planets orbit close to their host stars, high-energy irradiation can drive atmospheric escape, potentially removing large fractions of the atmosphere over geological time \citep{Luger_Barnes_2015, Lehmer_Catling_2017, Owen_2019}.

Photoevaporation has emerged as a key mechanism shaping exoplanet atmospheres \citep{Sanz-Forcada_2011,lopez-fortney-2013,Owen_2019}. Stellar X-ray and extreme ultraviolet (XUV) photons heat the upper atmosphere to overcome the gravitational bounding of the particles at the top of the atmosphere, leading to their escape to space \citep{Watson_1981, Lammer-2003}. This process may explain the observed radius valley \citep{Fulton_2017}, the lack of planets with radii between 1.5\,$R_\oplus$ and 2.0\,$R_\oplus$. In particular, hydrodynamic escape can drive the transition from a primary, low mean molecular weight hydrogen envelope to a secondary, high mean molecular weight atmosphere or even a bare rocky core \citep{lopez-fortney-2013,Kite-2020, Krissansen-Totton-2024}.
Core-powered mass loss \citep{Ginzburg_2018, Gupta-2019} is a complementary mechanism proposed to shape the evolution of the radius valley due to the planet’s cooling rocky core radiating heat, in addition to atmospheric escape. According to \citet{Owen-Schlichting-2024}, both mass loss mechanisms are dominant in different regimes: core-powered mass-loss is most significant for low-gravity planets (e.g., a few Earth masses or less) at high equilibrium temperatures ($\approx$\,500--2000\,K), while photoevaporation dominates for higher-gravity planets at lower temperatures.
On close-in rocky planets subject to intense stellar irradiation, atmospheric escape is most efficient at early times ($\sim10^6$--$10^8$\,yr) through XUV-driven hydrodynamic loss \citep{Luger_Barnes_2015, Owen_2019}, whereas escape driven by core-powered mass loss is supposed to operate over longer cooling timescales \citep[$\sim10^9\,$\,yr,][]{Gupta-2019,Johnstone-2019,Wordsworth_2022}.

Observational evidence of atmospheric escape has been reported in several exoplanetary systems through detections of neutral hydrogen Ly$\alpha$\,$\lambda$121.6\,nm and metastable helium He~{\sc i}\,$\lambda$1083.0\,nm absorption \citep{Vidal-Madjar-2004,Nortmann_2018,Spake_2021, Allart_2025}. Signatures of evaporating tails caused by intense atmospheric escape have been observed in gas giants \citep[HD\,209458\,b;][]{Vidal-Madjar-2004} and sub-Neptunes \citep[GJ\,436\,b;][]{Ehrenreich-2015}. Indirect evidence exists for smaller, terrestrial planets: Venus exhibits an enhanced D/H ratio compared to Earth \citep{Donahue_1982,Sossi_2020} and a hydrogen exosphere detected in Ly$\alpha$ \citep{Barth_1968}. Direct detections of atmospheric escape on Earth-sized exoplanets are still challenging \citep{Masson-2024}. Atmospheric escape can substantially alter the composition of rocky planets over $\sim10^6$--$10^9\,$\,yr. In this context, scenarios involving high mean molecular weight atmospheres have been explored for hot super-Earths orbiting M-dwarf stars. 
For GJ\,1132\,b, \citet{Schaefer_2016} suggested that a dense \element{CO_2} atmosphere may survive strong XUV irradiation. Newer observations suggest also a high mean molecular weight atmosphere scenario \citep{Diamond-Lowe-2018, Libby-Roberts-2022}, including a water-dominated atmosphere \citep{May-2023} or no atmosphere at all \citep{Xue-2024}. Several ultra-short-period planets appear to have lost their atmosphere very efficiently and end up as bare rock. LHS\,3844\,b does not have a thick atmosphere, consistent with a rapid and efficient loss of its primordial atmosphere \citep{Kreidberg_2019}, while models of L\,98--59\,d suggest that escape and interior-atmosphere evolution may have significantly altered its initial envelope \citep{Nicholls_2025_submitted}. Atmospheric evolution models of the TRAPPIST‑1 planets predict that close-in rocky worlds around an active M dwarf can lose a substantial fraction of their primordial atmospheres over gigayear timescales \citep{Bolmont_2017, Van_Looveren_2024, Thomas_2025}.
Modeling a planet's atmosphere and interior together during its evolution is essential, because the interior is the primary supplier and sink of volatiles: in- and outgassing together with redox state set the atmospheric composition and replenish (or sequester) gases on timescales that control atmospheric long-term retention \citep{Lichtenberg_2025, Nicholls_2025_submitted}. Observations of ultra-short-period planets such as TOI-561\,b \citep{teske_2025} suggest that atmospheres may be sustained by exchange with a molten interior reservoir, where volatiles are continuously replenished through outgassing from the magma ocean. This highlights that escape calculations neglecting interior replenishment may underestimate the number of planets that retain atmospheres.
Atmospheric escape itself therefore cannot be treated as an isolated sink: escape rates strongly shape the evolving volatile budget and interior thermal evolution, all of which feed back on each other \citep{Lichtenberg_2025}. 

Despite significant advances in observations and modeling, key questions remain unanswered regarding the role of hydrodynamic escape in shaping the atmospheric composition and evolution of molten rocky exoplanets. To address this, we developed a coupled interior-atmosphere model that includes hydrodynamic escape to simulate the evolution of magma oceans on rocky planets over geological timescales. We explored a broad parameter space to determine the conditions under which Earth-mass planets ($1\,M_\oplus$) and other rocky exoplanets can retain a secondary atmosphere or evolve into stripped, atmosphere-less bare rocks.

We introduce our coupled interior-atmosphere code, including atmospheric escape in Sect.\,\ref{methods section}. In Sect.\,\ref{results section}, we present our results exploring a large grid of parameters, including energy-limited escape on molten Earth-sized planets. In Sect.\,\ref{discussion section}, we discuss atmospheric timescales on magma ocean worlds and interaction with the elements trapped in the interior. We also explore the limitations of our modeling. We present our main conclusions in Sect.\,\ref{conclusions}.

\section{Methods}
\label{methods section}

\begin{figure*}
   \centering
    \includegraphics[width=\textwidth]{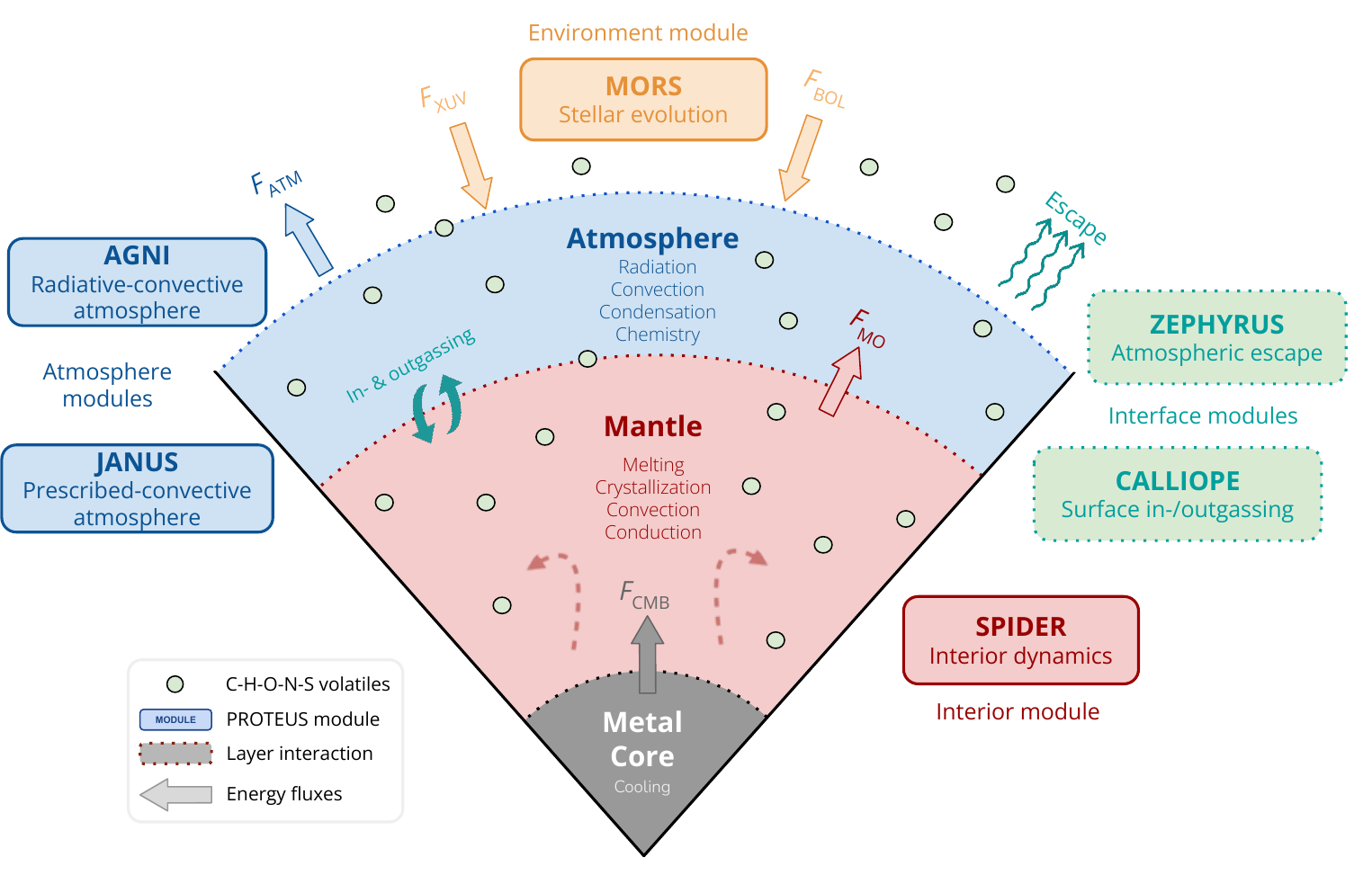}
    \caption{Schematic of the coupled interior-atmosphere evolutionary framework \texttt{PROTEUS}, showing the links between interior, surface, atmosphere, and stellar modules. Core components include \texttt{SPIDER} (mantle dynamics), \texttt{CALLIOPE} (surface in-/outgassing), \texttt{AGNI}/\texttt{JANUS} (radiative- and prescribed-convective atmospheres), \texttt{ZEPHYRUS} (atmospheric escape), and \texttt{MORS} (XUV stellar evolution). Modules exchange energy fluxes at each planetary interfaces---including core cooling, volatile outgassing, heat transport, heat loss to space, escape and stellar irradiation---while C--H--O--N--S volatiles are tracked throughout the mantle and atmosphere.}
    \label{proteus scheme}
\end{figure*}

\subsection{Coupled interior-atmosphere modeling framework}

We used the coupled one-dimensional interior-atmosphere framework \texttt{PROTEUS}\footnote{\url{https://proteus-framework.org}} \citep{Lichtenberg_2021_JGRP, Nicholls_2024_JGRP, Nicholls_2025_MNRAS}, which self-consistently simulates the evolution of rocky planets over gigayear timescales. The model couples distinct modules (see colored boxes in Fig.\,\ref{proteus scheme}) for mantle dynamics, volatile exchange, atmospheric structure, and stellar irradiation. We simulated the evolution of molten Earth-mass planets during the magma ocean era. In this work, we extended the framework by a self-consistent treatment of atmospheric escape for bulk composition via the new \texttt{ZEPHYRUS} module (Sect.\,\ref{section el escape methods}). 

\texttt{SPIDER} is a one-dimensional dynamic interior model tracking the thermal and rheological evolution of a planetary mantle (red box in Fig.\,\ref{proteus scheme}). It follows the transition from a fully molten state through a convecting magma ocean to full solidification. Following \citet{Salvador_2023}, \citeauthor{Nicholls_2024_JGRP} (\citeyear{Nicholls_2024_JGRP}; \citeyear{Nicholls_2025_submitted}), and \citet{Carone_2025}, we assumed a fully molten magma ocean at the start of the simulation with an initial specific entropy $S_0 = 3300\,\mathrm{J\,K^{-1}\,kg^{-1}}$. \texttt{SPIDER} provides depth-dependent profiles of specific entropy, temperature, melt fraction, and composition during the magma ocean evolution. This model includes heat transport from convection, phase mixing between solid and melt, separation, and gravitational settling, as well as core cooling represented by a lower boundary heat flux. Radiogenic heating and tidal dissipation were neglected, consistent with early magma ocean conditions where primordial heat dominates \citep{ElkinsTanton2008, Lichtenberg_2021_JGRP, Nicholls_2024_JGRP}; tidal effects are discussed in Sect.\,\ref{discussion model assumptions}. The mantle is modeled as a single silicate component (\element{MgSiO_3}) using the equation of state of \citet{WOLF2018}. We adopted melting curves that produced bottom-up crystallization, where solidification begins at the base of the mantle and progresses upward \citep{Bower_2019,Nicholls_2024_JGRP}. In this scenario, all volatiles are assumed to be released in the atmosphere during mantle solidification. As volatile trapping in the solid phase is neglected, our estimates for the outgassed volatile inventory represent an upper bound.

\texttt{CALLIOPE}\footnote{\url{https://proteus-framework.org/CALLIOPE/Explanations/model.html}} \citep{Bower_2022, SOSSI_2023, Nicholls_2024_JGRP, Nicholls_2025_MNRAS} computes exchange rates of in- and outgassed species between the interior and atmosphere, assuming a bulk silicate Earth mantle composition (green box in Fig.\,\ref{proteus scheme}). At each \texttt{PROTEUS} time step, volatile partitioning is recalculated using the magma temperature, surface pressure, and mantle melt fraction provided by \texttt{SPIDER}, yielding the bulk atmospheric composition supplied to the atmospheric module. Volatile partitioning is obtained assuming equilibrium chemistry and solubility laws \citep{DIXON_1995, ONEILL_2002, ARDIA_2013, ARMSTRONG2015, Dasgupta_2022, Gaillard_2022} for \element{H_2O}, \element{CO_2}, \element{N_2}, \element{S_2}, \element{SO_2}, \element{H_2S}, \element{NH_3}, \element{H_2}, \element{CH_4}, \element{CO}, and \element{O_2}. Following \citet{Bower_2022} and \citet{Gaillard_2022}, we assumed that surface pressure regulates volatile ingassing and outgassing, and that the vigorously convecting molten magma column is well mixed and remains in equilibrium with the atmosphere at each time step. After mantle solidification, outgassed compositions are fixed at their values at $T = 700$\,K. We assumed a thin conductive boundary layer at the magma ocean--atmosphere interface ($1\,\mathrm{cm}$ and thermal conductivity of $2.0\ \mathrm{W\,m^{-1}\,K^{-1}}$), which limits mantle-to-surface heat transport and modulates volatile exchange \citep{ElkinsTanton2008,Hamano-2015}.

Atmospheric structure is computed using the one-dimensional modules \texttt{JANUS} or \texttt{AGNI} \citep[blue boxes in Fig.\,\ref{proteus scheme};][]{Nicholls_agni_2025, Nicholls_2025_MNRAS}, which both generate pressure--temperature, radius, and composition profiles in equilibrium throughout the planet evolution. Both modules include condensation of volatiles and solve radiative transfer using the \texttt{SOCRATES} code \citep{edwards_studies_1996, amundsen_radiation_2014, Sergeev_2023}. It computes radiative fluxes on the specified $P$--$T$ grid using $k$-coefficients fitted to cross-sections derived from the DACE database \citep{Grimm-2021}, which incorporates ExoMol, HITRAN, and HITEMP linelist data. We accounted for Rayleigh scattering, but neglect the optical effects of aerosols. \texttt{JANUS} \citep{Graham-2021} constructs pressure--temperature profiles by assuming that the atmosphere initially follows a dry adiabat near the hot surface of the magma ocean. Then, it transitions to a moist pseudoadiabat where condensation occurs, and approaches an isothermal stratosphere at higher altitudes where the gas becomes optically thin \citep{Pierrehumbert-2010-book,Lichtenberg_2021_JGRP}. \texttt{AGNI} \citep{Nicholls_2025_tidal_l9869, Nicholls_2025_submitted, Nicholls_agni_2025,Nicholls_2025_MNRAS} generates $P$--$T$ profiles while accounting for the possibility of convective stability in the atmosphere. The atmosphere is solved with an energy-conserving profile formalism, ensuring that energy fluxes across each layer are conserved. As presented by \citet{Nicholls_2025_submitted}, \texttt{AGNI} uses non-ideal gas equations of state (EOS) for the all volatiles species considered in \texttt{CALLIOPE}, except for \element{S_2} which is treated as an ideal gas. \texttt{JANUS} employs prescribed adiabatic pressure--temperature profiles, whereas \texttt{AGNI} computes self-consistent radiative--convective atmospheres. For both atmospheric modules, $P$--$T$ profiles are solved on a pressure grid extending from the surface pressure $P_{\mathrm surf}$ (provided by the outgassing module) up to a top-of-atmosphere pressure set at $P_\mathrm{top} = 10^{-5}\,\mathrm{bar}$. At each time step, the radiative balance accounts for both the incoming stellar irradiation and the thermal emission from the magma ocean surface \citep{Lichtenberg_2021_JGRP}. We adopted a surface albedo of 0.1, following experimental indications of low reflectivity of magma ocean surfaces \citep{Essack-2020}.

\begin{figure}
    \resizebox{\hsize}{!}{\includegraphics{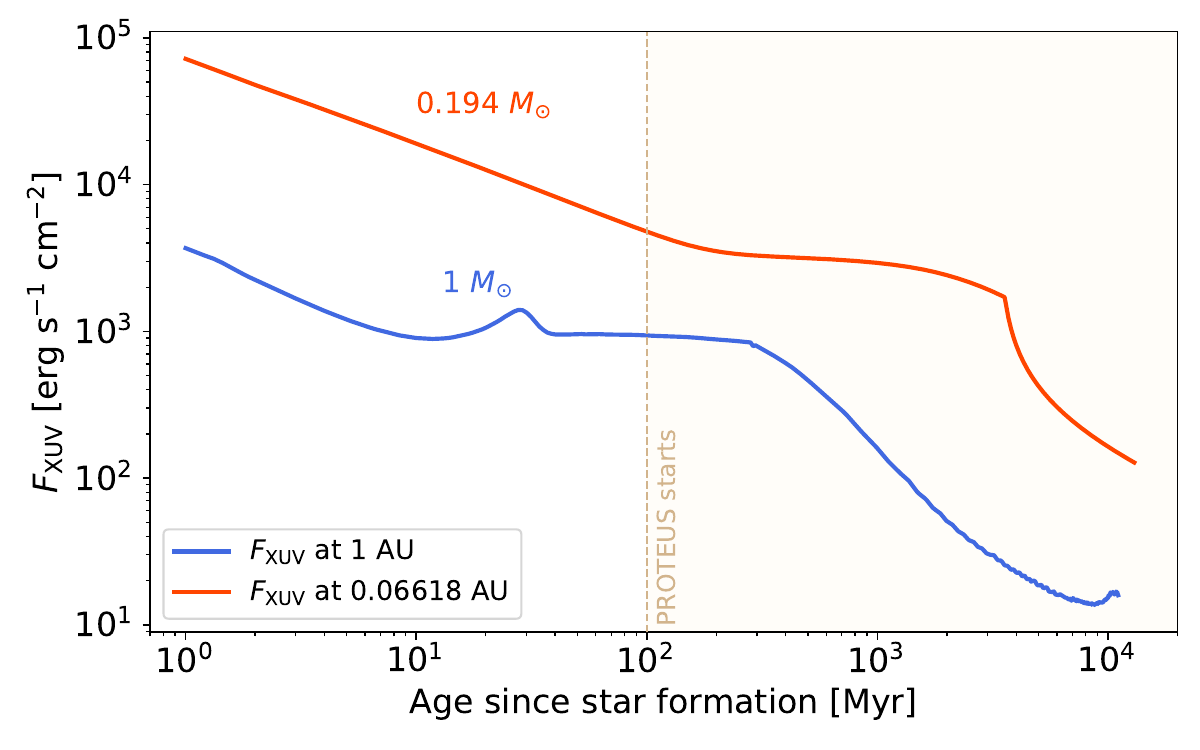}}
    \caption{XUV flux received by a planet orbiting a G star (blue line) and an M dwarf (orange line) since stellar birth. Fluxes are computed at orbital separations corresponding to present-day Earth bolometric instellation (1\,au for $1\,M_\odot$ and 0.06618\,au for 0.194\,$M_\odot$). \texttt{PROTEUS} simulations start at 100\,Myr (vertical dashed line).}
    \label{plot xuv evolution stellar tracks}
\end{figure}

We modeled the stellar evolution of a solar-type star and an M dwarf using the stellar evolution tracks from \citet{2013_spada} implemented in \texttt{MORS} \citep[orange box in Fig.\,\ref{proteus scheme} --][]{Johnstone_2020, Johnstone_2021}. For the $1\,M_\odot$ solar case, we adopted the solar spectrum \citep{gueymard_sun_2004} and use a median-rotation track at the 50th percentile compared to the stellar population at the same mass. For the M-dwarf simulations, we also considered a medium rotator of spectral type M4 with a mass of 0.194\,$M_\odot$, using the GJ\,1132 spectrum from the MUSCLES database \citep{Wilson-2025}. All simulations presented in this work start at 100\,Myr after star formation, after typical protoplanetary disk dispersal \citep[$10$--$15\,$\,Myr,][]{Haisch-2001, Williams-2011,Ribas-2015-protoplanetarydisks}.
At each  \texttt{PROTEUS} time step, \texttt{MORS} provides the X-ray, UV, and bolometric band luminosities of the star. The bolometric flux $F_{\star}$, is updated at small time steps (every 100\,years) to capture shorter-term fluctuations. The full stellar spectrum used in the \texttt{SOCRATES} radiative transfer scheme is updated every 100\,Myr.
Figure\,\ref{plot xuv evolution stellar tracks} shows stellar XUV radiation \citep[$0.1$--$91\,\mathrm{nm}$, as defined by][]{Johnstone_2021} evolution for different stellar types. The received XUV flux differs by about an order of magnitude between a G star (blue line) and an M dwarf (orange line). The M dwarf exhibits a much longer saturation phase ($\sim3$\,Gyr) compared to the G star ($\approx300$\,Myr); and higher XUV fluxes at fixed instellation distance (e.g. at 1\,Gyr: $F_{\mathrm{XUV,M}} = 4\times10^{3}\ \mathrm{erg\,s^{-1}\,cm^{-2}}$ versus $F_{\mathrm{XUV,G}} = 2\times10^{2}\ \mathrm{erg\,s^{-1}\,cm^{-2}}$).
XUV radiation is the main driver of hydrodynamic atmospheric escape during the saturated phase, implying stronger atmospheric loss for stars with longer saturation timescales \citep{lopez-fortney-2013, Owen_2019}. However, XUV luminosities of young stars remain uncertain due to ISM absorption and stellar rotation history \citep{Ribas_2005, Tu-2015, Richey-Yowell-2019-hazmat, Peacock_2020}; these uncertainties are discussed in Sect.~\ref{section fxuv uncertainties}.

\subsubsection{Hydrodynamic escape}
\label{section el escape methods}

We computed bulk atmospheric escape using the \texttt{ZEPHYRUS} module (green box in Fig.\,\ref{proteus scheme}), which implements an energy-limited (EL) formalism, coupled to the stellar evolution model from \texttt{MORS}. We assumed an hydrodynamic escape, which is expected to dominate for close-in exoplanets during the early stages of their evolution, when stellar XUV emission is high \citep{Watson_1981, Lammer-2003, Ribas_2005}. In the EL regime, stellar XUV radiation absorbed at the top of the atmosphere heats the gas, allowing it to expand and escape once it overcomes the planet’s gravity. Following \citet{Watson_1981}, \citet{Erkaev_2007}, and \citet{lopez-fortney-2013} the mass loss rate for energy-limited escape is given by 
\begin{equation}
    \dot M_\mathrm{EL}= \epsilon \pi \frac{R_\mathrm{XUV}^3 F_\mathrm{XUV}}{G M_\mathrm{p} K_\mathrm{tide}}
    \label{equation EL escape}
\end{equation}
with $\epsilon$ the escape efficiency factor, $R_\mathrm{XUV}$ the radius down to which XUV radiation penetrates the atmosphere in m (i.e the altitude at which escape starts), $F_\mathrm{XUV}$ the  stellar XUV flux received by the planet in $\mathrm{W\,m^{-2}}$, G the gravitational constant in $\mathrm{N\,m^{2}\,kg^{-2}}$, $M_\mathrm{p}$ the mass of the planet in kg and $K_\mathrm{tide}$ the tidal correction factor (here set to 1 for no tidal effects).

The mass-loss efficiency factor for hydrodynamic escape, $\epsilon$, quantifies the proportion of the incoming $F_\mathrm{XUV}$ that is effectively transformed to do work on the gas for atmospheric escape. This factor typically ranges from 0.1 to 0.6 \citep{Owen_Wu_2013, Luger_Barnes_2015, Lehmer_Catling_2017, Wordsworth_2018}---although some studies have adopted values up to 1.0 \citep{Lecavelier-Des-Etangs-2007, Ehrenreich-2011}. In practice, $\epsilon$ is expected to vary with planet size, atmospheric composition, and dominant cooling mechanisms operating in the upper atmosphere. Atomic line cooling, molecular emission, and ionization losses can significantly divert XUV energy away from heating the bulk gas, thereby reducing the effective energy available for hydrodynamic escape \citep{Shematovich-2014, Nakayama-2022,Chatterjee_2024,Yoshida-2024}. In this work, we explored $\epsilon$ values from 0.1 to 1.0 to assess the impact of varying efficiency on escape rates. We discuss the physical plausibility and implications of these choices in Sect.\,\ref{discussion epsilon section}.

The escape radius $R_\mathrm{XUV}$ denotes the altitude at which the atmosphere becomes optically thick to stellar XUV photons (often approximated as the exobase). Recent studies \citep{Luger_Barnes_2015,Moore_2023} assume $R_\mathrm{XUV} = R_\mathrm{p}$, which provides a conservative (lower) bound on the estimated hydrodynamic mass-loss rate. Allowing $R_\mathrm{XUV} > R_\mathrm{p}$ increases the effective XUV absorbing area, thus a larger volatile inventory is available for escape, resulting in higher escape rates.
\citet{Salz-2015} and \citet{Chen-2016} proposed semi-empirical prescriptions to estimate $R_\mathrm{XUV}$ for large gas planets surrounded by \element{H_2}/\element{He} envelopes. But these formulations were calibrated for giant planets, their applicability to compact rocky planets with high mean molecular weight atmospheres remains uncertain. In our implementation with \texttt{PROTEUS}, $R_\mathrm{XUV}$ is set via an input pressure level $P_\mathrm{XUV}$ (e.g., the pressure where XUV absorption becomes significant) and is updated as the atmospheric structure changes. For example, some studies define $R_\mathrm{XUV}$ at a fixed reference pressure \citep[e.g., $20\,\mathrm{mbar}$ in][]{Baumeister_2023}. In this work, $P_\mathrm{XUV}$ is specified as an input parameter and $R_\mathrm{XUV}$ is recomputed at each time step accordingly.

In this work, we considered a wide range of atmospheric compositions, from hydrogen-dominated to high mean molecular weight mixtures, where fractionation may occur. We assumed that all species in the atmosphere are entrained in the hydrodynamic outflow in bulk, neglecting fractionation for simplicity. As in \citet{Nicholls_2025_submitted}, we assumed that the escaping gas has the same elemental proportions (by mass) as the outgassed species calculated by \texttt{CALLIOPE}. At each time, we computed the total escape rate (Equation\,\ref{equation EL escape}) and distribute it among species according to their atmospheric mass mixing ratios. Hence, the escape flux [in $\mathrm{kg\,s^{-1}}$] of each element is proportional to its abundance in the atmosphere. Importantly, although our treatment of hydrodynamic escape itself is non-fractionating, escape can act to fractionate the planet's bulk volatile inventory because each element's loss rate is linked to its outgassed proportion rather than the bulk planetary composition. 

\subsection{Grid configuration} 
\label{grid section}

\begin{table*}
\caption{Parameter space tested in our grid of evolutionary models.}
\label{table grid parameters}
\centering
\begin{tabular}{lll}
\hline\hline
Parameter [unit] & Symbol & Values \\
\hline
Stellar mass [$M_\odot$] & $M_*$ & 0.194, 1.0 \\
Atmosphere treatment & ... & \texttt{AGNI} (radiation--convection), \texttt{JANUS} (prescribed--convection) \\
Semimajor axis [au]\tablefootmark{a} & $a$ & 0.1, 0.5, 1.0 ($M_* = 1.0\,M_\odot$) \\
& 
& 0.006618, 0.033091, 0.066182 ($M_* = 0.194\,M_\odot$) \\
XUV pressure [bar] & $P_{\mathrm{XUV}}$ & $10^{-5}$, $10^{1}$ \\
Escape efficiency factor [non-dim.] & $\epsilon$ & 0.1, 0.5, 1.0 \\
Oxidation state [$\Delta$IW]\tablefootmark{b} & $f\mathrm{O}_2$ & $-4$, $0$, $+4$ \\
C/H ratio [MMR]\tablefootmark{c} & C/H & 0.1, 1.0, 2.0 \\
Total H inventory [Earth oceans]\tablefootmark{d} & [H] & 1, 5, 10 \\
\hline
\end{tabular}
\tablefoot{
\tablefootmark{a} {Semimajor axes depend on stellar mass: for a solar-type star ($M_* = 1.0\,M_\odot$) the reference distances are 0.1, 0.5, and 1.0\,au; for an M dwarf ($M_* = 0.194\,M_\odot$), the corresponding values are 0.006618, 0.033091, and 0.066182\,au, scaled to match Earth's instellation today \citep[$\mathcal{F}_{\oplus}=4.64\,\mathrm{erg\,s^{-1}\,cm^{-2}}$;][]{Ribas_2005}.} \quad
\tablefootmark{b} {Log-units relative to the iron--wüstite (IW) buffer \citep{ONEILL_2002}. $f\mathrm{O}_2$ is in units of bar.} \quad
\tablefootmark{c} {Mass mixing ratio in the mantle and atmosphere combined.} \quad
\tablefootmark{d} {Mass of 1 Earth ocean $= 1.4\times10^{21}\,\mathrm{kg}$ of \element{H_2O} \citep{Genda_2008}.}
}
\end{table*}

We investigated the effect of hydrodynamic bulk escape on Earth-like exoplanets coupled to the interior-atmosphere evolution framework \texttt{PROTEUS} up to gigayear timescales. Our simulation grid explored a broad range of parameters described in Table\,\ref{table grid parameters}.  We ran two grids of 972 simulations each for molten Earth-mass planets orbiting a Sun-like star and an M dwarf. The grids were run concurrently on the H\'abr\'ok\footnote{\url{https://wiki.hpc.rug.nl/habrok/start}} high performance computing cluster, with up to 500 jobs executing in parallel, requiring approximately 4--5 days of wall-clock time.

All planets were placed on circular orbits ($e$ = 0) around their star. We neglected tidal heating effects, which for eccentric orbits can prolong magma ocean lifetimes \citep{Farhat2025ApJ,Nicholls_2025_tidal_l9869}; this simplification yields to shorter solidification timescales. To treat the whole-planet climate using a representative one-dimensional column calculation, the zenith angle for incoming stellar radiation is set to $\theta_\mathrm{zenith}= 48.19^\circ$ and a scaling factor $s_0 = 0.375$ is applied to the computed stellar spectrum \citep{Cronin-2014, Hamano-2015, Nicholls_2024_JGRP}. These parameters are appropriate for a planet which is asynchronously rotating, although, in practice, their particular values have little impact on the climate within a column model. The planet's structure is set by a $1.0\,M_\oplus$ total mass (including silicate mantle, metallic core, and volatiles). The core radius is fixed at 55\% of the interior radius, consistent with previous studies \citep{Nicholls_2025_submitted} and Earth’s core size \citep{Dziewonski-1981}. 
The core density is set to $10\,738.33\ \mathrm{kg\,m^{-3}}$ and the specific heat capacity to $880\ \mathrm{J\,K^{-1}\,kg^{-1}}$, matching Earth-like values \citep{Dziewonski-1981} and consistent with values employed in recent magma ocean evolution models \citep{Bower_2018,Nicholls_2024_JGRP, Nicholls_2025_submitted}. For volatile inventories, we set the initial mantle nitrogen concentration to 2.0\,ppmw relative to the mantle mass, consistent with Earth's values \citep{Johnson-Goldblatt-2015, Li-2024} as well as sulfur, set to 200.0\,ppmw \citep{Sun-2020}. We noted that although most of Earth’s nitrogen is now in the atmosphere, our adopted mantle \element{N} and \element{S} abundances provide a reference; additional volatiles (\element{H}, \element{C} and \element{O}) are varied across the parameter grid. 

In this work, planets orbiting a Sun-like star ($1\,M_\odot$) at 0.1\,au reach equilibrium temperatures of $T_{\rm eq}\approx$\,910\,K, while at 0.5\,au and 1.0\,au, $T_{\rm eq}$ drops to $\approx$\,405\,K and $\approx\,290$\,K, respectively. In the low-mass star regime ($0.194\,M_\odot$), planets receiving the same instellation as in the solar-mass case reach equilibrium temperatures of 
$T_{\rm eq} \approx 1100$\,K at $0.00662\,$\,au, $\approx\,490$\,K at $0.03309\,$\,au, and $\approx\,340$\,K at $0.06619\,$\,au. 
The equilibrium temperatures we derived ($\approx290$--$1100$\,K, assuming a zero Bond albedo) cover the range of currently observed rocky exoplanets, including temperate planets such as TRAPPIST‑1\,e \citep[$T_{\rm eq}\approx\,250$\,K;][]{Gillon_2017}, planets with moderate heating such as L\,98‑59\,d with $T_{\rm eq}\approx\,416$\,K \citep{Demangeon_2021} or TRAPPIST‑1\,b \citep[$T_{\rm eq}\approx\,400$\,K;][]{Gillon_2017} and hotter planets, such as GJ\,1132\,b \citep[$T_{\rm eq}\approx\,584$\,K;][]{Xue-2024} and GJ\,486\,b \citep[$T_{\rm eq}\approx\,706$\,K;][]{Caballero_2022}. Ultra-hot super-Earths, such as 55\,Cancri\,e \citep[$T_{\rm eq}\approx\,2500$\,K;][]{Hu_2024} and TOI‑561\,b \citep[$T_{\rm eq}\approx\,2310$\,K;][]{teske_2025}, lie beyond the range considered here but are likely in a permanent magma ocean stage.

In this study, a simulation ends when one of the following criteria is met:

\begin{itemize}
    \item Complete solidification: The mantle melt fraction falls below the critical value $\Phi_{\rm crit} < 0.005 $, below which we assumed the mantle is fully solid \citep{Miller-2014, Laumonier-2017} and end the magma ocean stage.
    \item Atmosphere loss: The surface pressure $P_\mathrm{surf}$ drops below $P_\mathrm{stop} = 1$\,bar, at which point we considered that the atmosphere has been effectively removed to space. Any residual gas at this level is thin, provides negligible greenhouse warming, and does not significantly affect magma‑ocean cooling \citep{Wordsworth_2022, Salvador_2023}. Secondary atmospheres below this value are very difficult to detect with current observatories.
    \item Radiative equilibrium: The planet reaches global energy balance, meaning that the difference between planetary outgoing and incoming fluxes is less than $0.2\,\mathrm{W\,m^{-2}}$ between two successive time steps. The simulation finishes in a quasi-steady-state regime. This threshold is more stringent than the  $0.8\,\mathrm{W\,m^{-2}}$ criterion adopted by \citet{Nicholls_2024_JGRP} and
    and corresponds to a negligible residual energy imbalance relative to the planetary cooling flux.
\end{itemize}

Out of the 1944 initial simulations from our parameter space (972 for a Sun-like star and 972 for the M dwarf), 383 reached steady-state with a balanced energy budget around a Sun-like star (117 for planets orbiting an M dwarf), 294 fully solidified around a G star (against 372 for the $0.194\,M_{\odot}$ star grid) and 154 lost their atmospheres entirely in a G star system (against 354 in the M star case). The results presented here are based on the numerically converged simulations, which represent 86\% of the original simulation grid; the remaining cases did not reach numerical convergence within the adopted stopping criteria. Planets in radiative equilibrium can remain in a prolonged magma ocean phase, as their energy balance prevents rapid solidification, provided the atmosphere is retained. Even if a planet loses its atmosphere while still molten, the absence of atmospheric insulation prevents the magma ocean from being sustained over long timescales. We did not simulate planetary evolution beyond the point of magma ocean solidification or after complete atmosphere loss; the subsequent cooling, solid-state mantle tectonics, weathering processes, and volcanic degassing are beyond the scope of this work.

\section{Results}
\label{results section}

\subsection{Influence of hydrodynamic escape on atmospheric evolution}

\begin{figure*}
    \centering
    \includegraphics[width=\textwidth]{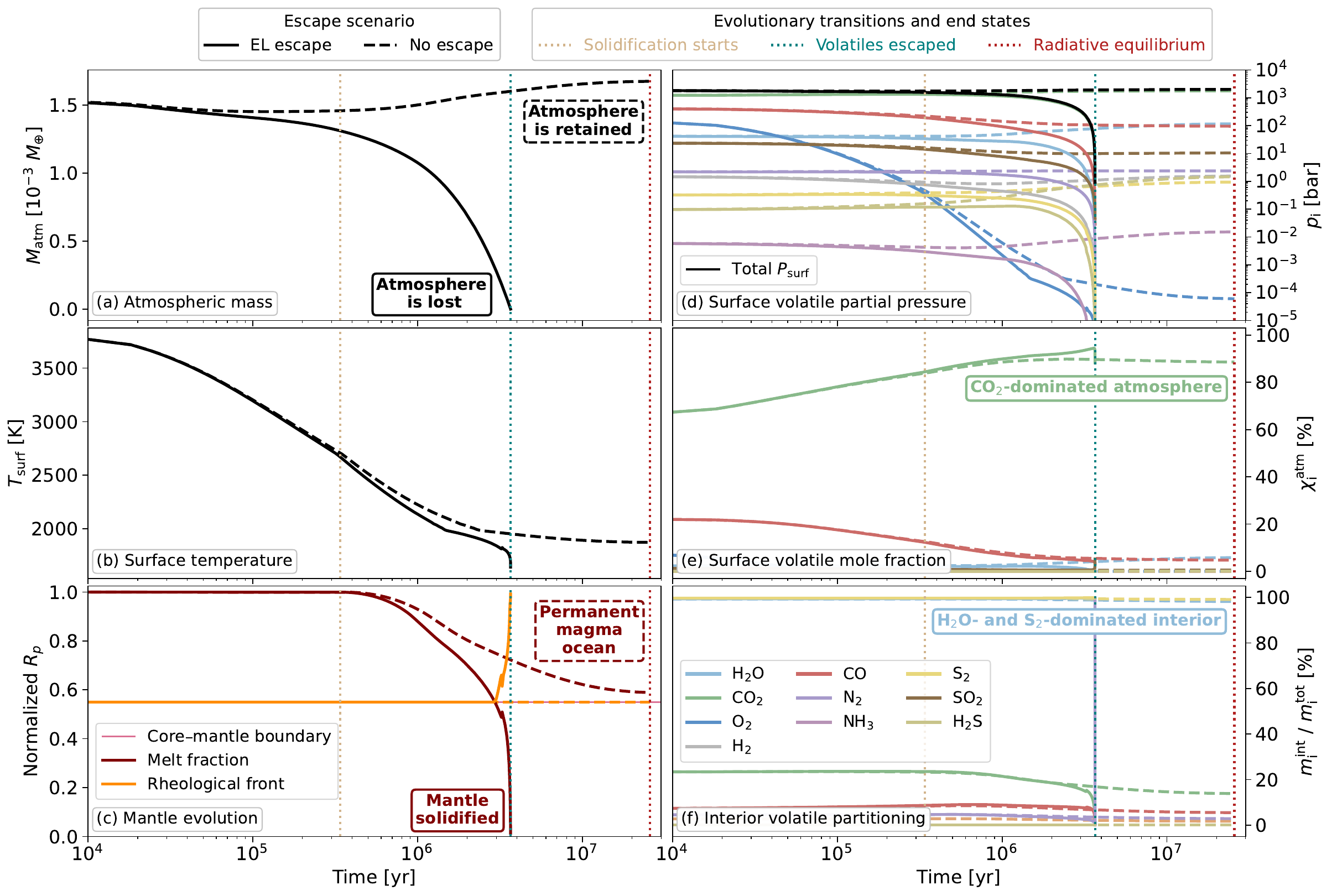}
    \caption{Time evolution of an Earth-size planet orbiting a Sun-like star at 0.1\,au, including hydrodynamic escape (solid lines) or not (dashed lines). With hydrodynamic escape, all outgassed volatile species are lost within 3.6\,Myr (dotted cyan lines), whereas without escape, the planet remains in a magma ocean state, reaching radiative equilibrium after 25.6\,Myr (dotted red lines).}
    \label{evolution escape on off}
\end{figure*}

Figure\,\ref{evolution escape on off} shows the time evolution of an Earth-analog planet ($1\,M_{\oplus}$) orbiting a Sun-like star at 0.1\,au, simulated with (solid lines) and without (dashed lines) hydrodynamic atmospheric escape. The planet is initialized with a volatile inventory of 10\,Earth oceans of hydrogen and a carbon content defined by a C/H ratio of 2.0 (i.e., the ratio of total carbon mass to total hydrogen mass in the mantle--atmosphere system). Both simulations assume an oxidized mantle ($f$O$_2$\,=\,IW+4). In the hydrodynamic escape scenario, we assumed an escape efficiency factor of $\epsilon = 1.0$ corresponding to fully efficient escape and an initial XUV pressure of $P_\mathrm{XUV} = 10$\,bar. Atmospheric structure was resolved using the radiative-convective model \texttt{AGNI} \citep{nicholls_beyond_2026}.

In the scenario with strong hydrodynamic escape, an initial atmospheric mass of $1.5 \times 10^{-3} M_\oplus$ (panel b), corresponding to 0.15\% of the planetary mass---roughly 1800 times larger than Earth’s present atmospheric fraction---is removed within 3.6\,Myr (dark cyan dotted lines), thereby inducing rapid mantle solidification. The simulation was terminated once the planet’s atmosphere was considered effectively lost, defined by a surface pressure $P_\mathrm{surf} < 1$\,bar. The mass-loss rate remains mostly constant over time, $\approx 5 \times 10^{10}\,\mathrm{g\,s^{-1}}$. The efficiency of atmospheric escape in this case is driven by the high escape efficiency factor ($\epsilon = 1.0$) and the planet's close proximity to its host star (0.1\,au), resulting in an incident XUV flux, $F_\mathrm{XUV}$, four orders of magnitude higher than present-day Earth \citep[$\mathcal{F}_{\oplus} = 4.64\,\mathrm{erg\,s^{-1}\,cm^{-2}}$;][]{Ribas_2005}.

In contrast, the no-escape simulation reaches radiative equilibrium after 25.6\,Myr (dark red dotted lines), staying in a permanent magma ocean state. By definition, in the no-escape scenario, the volatile fraction remains intact after 25.6\,Myr. However, the atmosphere builds up over time and strongly changes composition even if no escape is included. This indicates that energy-limited hydrodynamic escape is capable of depleting a massive planetary atmosphere over a timescale of a few million years, thereby strongly influencing both the interior evolution and the planet’s volatile content. 

\subsubsection{Magma ocean solidification timescale with atmospheric loss}

In both the escape and no-escape scenarios, the mantle begins to solidify at 0.3\,Myr, as indicated by a melt fraction dropping below 1 (dotted beige lines).
In the scenario including hydrodynamic escape, the rapid loss of the \element{CO_2}-dominated atmosphere (panel e) removes greenhouse insulation, shown by a sharp decrease in surface pressure ($P_\mathrm{surf}$, panel d) and a corresponding drop in surface temperature ($T_\mathrm{surf}$, panel b). The temperature decline corresponds to mantle solidification (panel c). During the final Myr of the simulation, atmospheric escape leads to a progressive thinning of the atmosphere (panel d), thereby accelerating the cooling of the magma ocean. Once the atmosphere is entirely depleted, the mantle undergoes rapid solidification starting at 3\,Myr.

In contrast, when atmospheric escape is neglected, the mantle remains in a permanent magma ocean state at the end of the simulation ($25.6 \times 10^6$\,yr). Panel c shows that the melt fraction decreases slowly over time but does not reach the core-mantle boundary at 0.55\,$R_{\oplus}$, preventing complete crystallization. The planet thus reaches a radiative equilibrium state, maintaining both a persistent magma ocean and a thick atmosphere for at least 25.6\,Myr. This indicates that the magma ocean period is significantly prolonged when atmospheric escape is ignored.

Accounting for hydrodynamic escape can thus drastically shorten the magma ocean crystallization timescale: from over 25.6\,Myr without considering escape to 3.6\,Myr with escape. This difference highlights the crucial role of atmospheric loss in regulating planetary interior and climate evolution.

\subsubsection{Evolution of volatile inventory under escape}

Panel d and panel e of Fig.\,\ref{evolution escape on off} show the evolution of volatile species exchanged between the atmosphere and magma ocean through outgassing and ingassing, while panel f shows the volatile inventory remaining dissolved in the magma ocean for both escape and no-escape scenarios. At $10^6$\,yr, the atmosphere in both cases is dominated by \element{CO_2} ($\approx\,90\%$; green line in panel f), with a smaller fraction of \element{CO} ($\approx\,6\%$; red line) and trace amounts (< 1\%) of \element{H_2O} and \element{SO_2}.

In the escape model, the atmosphere remains \element{CO_2}-dominated throughout the planet’s evolution (panel e) until substantial atmospheric escape occurs at $3.5\times10^{6}$\,yr. After the start of solidification (0.3\,Myr), \element{CO_2} stored in the magma ocean is progressively outgassed (green line, panel f), replenishing the atmosphere as escape depletes it. The \element{CO_2} molar atmospheric abundance increases from 85\% at the start of solidification, to about 95\% shortly before the atmosphere is lost. Meanwhile, the abundance of \element{CO} steadily decreases as the planet cools, being thermochemically favored at higher temperatures ($T>$\,1500\,K), whereas \element{CO_2} becomes increasingly favored at lower temperatures \citep{Schaefer_2010,Liggins_2023}. Near the end of the simulation, rapid crystallization from the base of the mantle to the surface releases a minor fraction of \element{SO_2} (5\%; brown line, panel e). Late-stage mantle enrichment of \element{N_2} (dark purple line), rising from a few percent to nearly the total interior volatile content, reflects outgassing during the final stages of mantle solidification. Thus, atmospheric escape strongly shapes the evolution of planetary atmospheres by controlling both the size of the atmospheric volatile reservoir and its long-term stability.

In contrast, in the no-escape model, surface volatile partial pressures slightly increase compared to their initial values due to replenishment by outgassing. \element{H_2O} rises from 41\,bar to 116\,bar, \element{NH_3} from $6 \times 10^{-3}$\,bar to $2 \times 10^{-2}$\,bar, and \element{S_2} from 0.32\,bar to 0.94\,bar. The total surface pressure remains nearly constant, rising from $P_{\mathrm{surf}, t_0} = 1814$\,bar to $P_{\mathrm{surf}, t_\mathrm{end}} = 1997$\,bar, as the \element{CO_2}-dominated atmosphere is primarily outgassed from the start. In the no escape scenario, the total atmospheric mass (panel a) gradually increases due to continuous volatiles degassing which is not counterbalanced by atmospheric loss. The atmosphere remains \element{CO_2}-dominated until the end of the simulation (25.6\,Myr), sustained in equilibrium with the mantle (panel f). The \element{CO} abundance decreases with time (as lower temperatures make carbon monoxide less stable), allowing \element{H_2O} to become the second most abundant atmospheric species, $\approx\,6\%$ (see panel e) once the planet reaches radiative equilibrium. Because of its high solubility in silicate melts \citep{Kite-2021,Lichtenberg_2021_JGRP,SOSSI_2023}, water is outgassed later than \element{CO_2} promoting the formation of a secondary, water-rich atmosphere. Thus, in the absence of atmospheric escape, volatiles progressively accumulate, replenishing the atmosphere and altering its composition over geological timescales.

\begin{figure*}
    \centering
    \begin{minipage}[b]{0.49\textwidth}
        \centering
        \includegraphics[width=\textwidth]{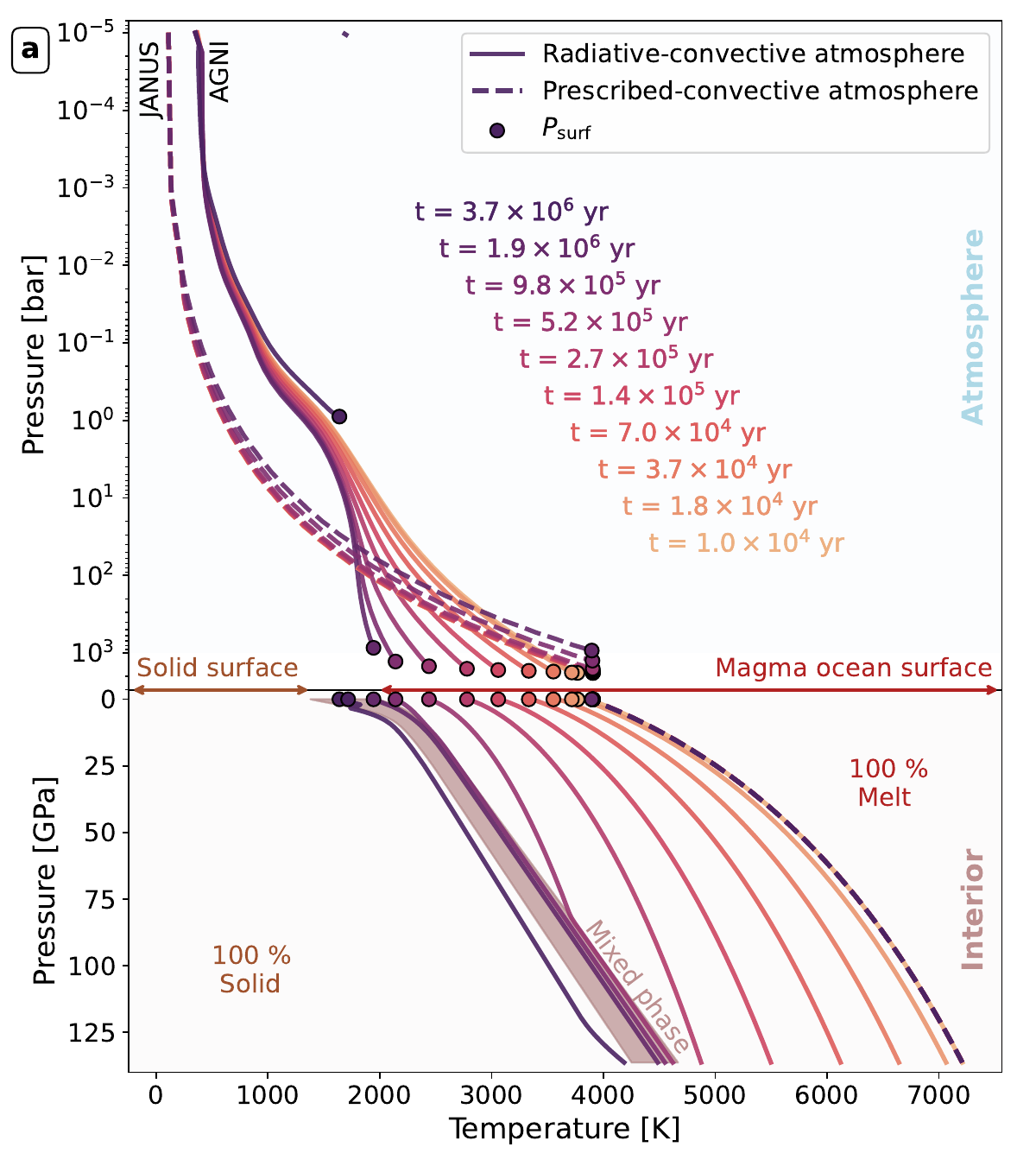}
    \end{minipage}
    \hfill
    \begin{minipage}[b]{0.49\textwidth}
        \centering
        \includegraphics[width=\textwidth]{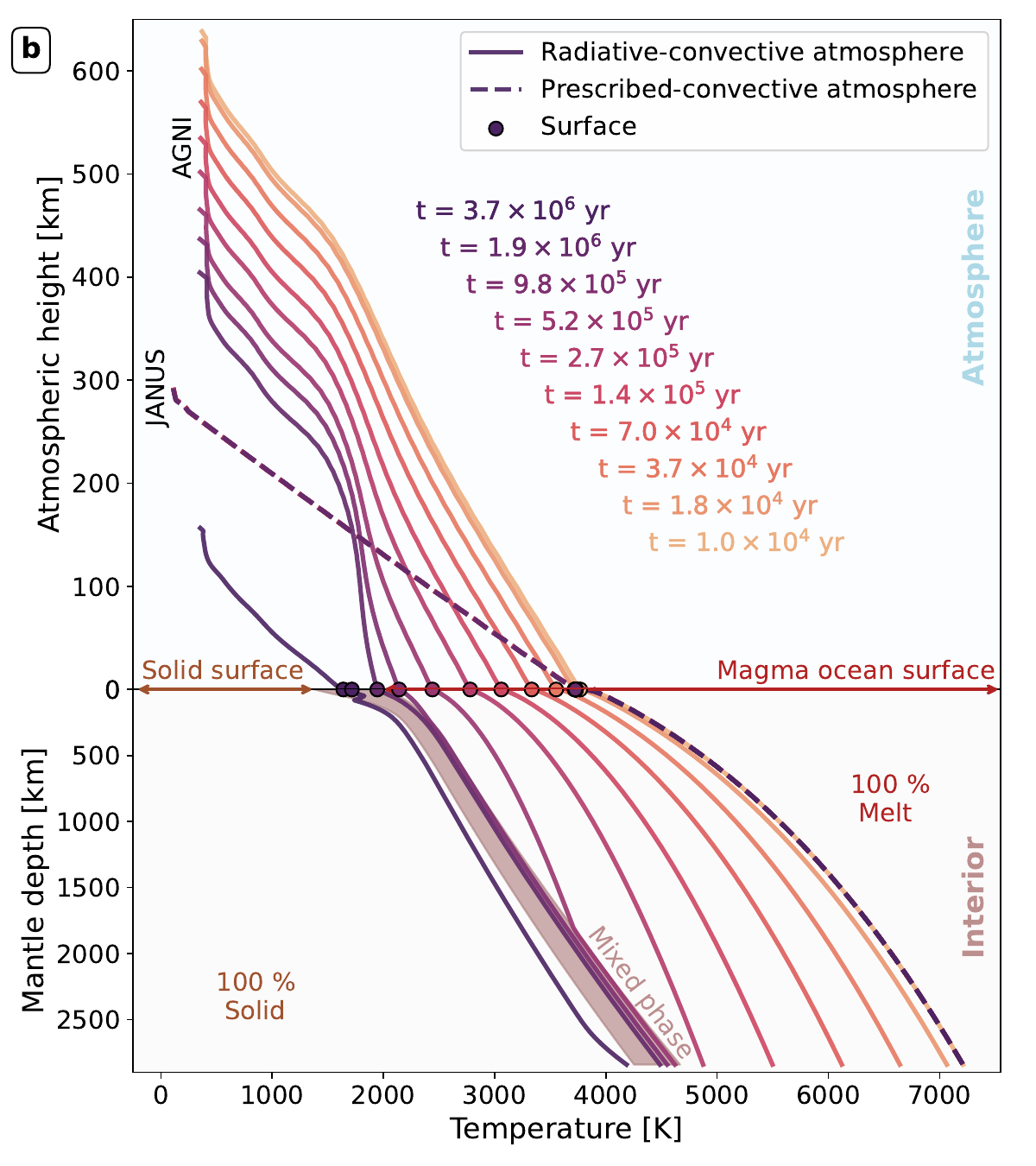}
    \end{minipage}
    \caption{Pressure--temperature and thermal structure evolution for a planet
    undergoing hydrodynamic escape, comparing a prescribed-convective
    model (\texttt{JANUS}, dashed lines) with a radiative--convective
    atmospheric treatment (\texttt{AGNI}, solid lines).
    Panel (a): Pressure--temperature profiles of the atmosphere and
    interior.
    Panel (b): Evolution of the thermal and vertical structure of the
    atmosphere and interior for the same planet as in
    Fig.\,\ref{evolution escape on off}.}
    \label{TP_profiles_escape_ON_AGNI_JANUS}
\end{figure*}

\subsection{Impact of radiative-convective atmospheres on escape and evolution}
\label{section agni janus difference}

Figure\,\ref{TP_profiles_escape_ON_AGNI_JANUS} presents pressure-temperature (panel a, left) and height-temperature (panel b, right) profiles at times sampled from our simulated evolution of an Earth-sized planet orbiting a Sun-like star at 0.1\,au (as in Fig.\,\ref{evolution escape on off}). We included atmospheric escape and also considered two different atmosphere treatments: prescribed-convective (\texttt{JANUS}, dashed lines) and fully radiative-convective (\texttt{AGNI}, solid lines).

Allowing for convective shutdown---via the \texttt{AGNI} models---causes the deeper atmospheric layers to enter a radiative regime, leading to gradual planetary cooling over time. A radiative-convective treatment of the atmosphere thereby strongly influences the thermal interior evolution of the planet: as the atmosphere cools near the surface, the temperature of the underlying magma ocean decreases accordingly. Decreasing temperatures reduce the scale height throughout the simulation. However, at the same time, atmospheric loss depletes the atmosphere, leading to substantial decreases in atmospheric height toward the end of the simulation using \texttt{AGNI} (panel b, solid lines). After 1\,Myr of evolution, the mantle enters a mushy regime, crystallizing from the bottom up. Simultaneously, atmospheric escape efficiently removes material, causing surface pressures to decrease (circular points in panel a). By the end of the simulation (3.6\,Myr), the mantle is almost fully solidified in the \texttt{AGNI} case and the surface pressure has dropped from 1814\,bar to 0.89\,bar due to near-complete atmospheric loss (panel a).

In contrast, simulations that consider only convective layers for atmospheric energy transport (\texttt{JANUS}, dashed lines) show little evolution from the initial state: the surface pressure remains near $P_\mathrm{surf}\approx 10^{3}$\,bar with a thick \element{CO_2} atmosphere until 3.5\,Myr. The atmospheric scale height is approximately constant over time (panel b) and the mantle remains fully molten in a persistent magma ocean stage. In this case, the convective treatment of the atmosphere maintains the planet in a quasi-steady state over long timescales ($3.5\times10^{6}$\,yr). This behavior, in the absence of escape, is consistent with previous modeling \citep{Nicholls_2025_MNRAS}. However, during the final $0.1\,$\,Myr of the simulation, the atmosphere is completely lost ($P_\mathrm{surf}=0$), leaving only a tenuous upper-boundary remnant at $P\sim10^{-5}$\,bar (faint purple feature), while the scale height collapses from $\approx\,292\,\mathrm{km}$ down to $\approx 5\,\mathrm{km}$ (unresolved in the figure due to scale). 

Overall, these atmosphere profiles demonstrate that---while a purely convective atmosphere maintains a long-lived magma ocean---the inclusion of radiative layers allows more efficient atmospheric cooling. Together with ongoing atmospheric loss, these layers accelerate magma ocean solidification, which highlights the combined importance of radiative processes and atmospheric escape in shaping planetary evolution.

\subsection{Parameter space exploration}

\begin{figure*}
   \centering
    \includegraphics[width=\textwidth]{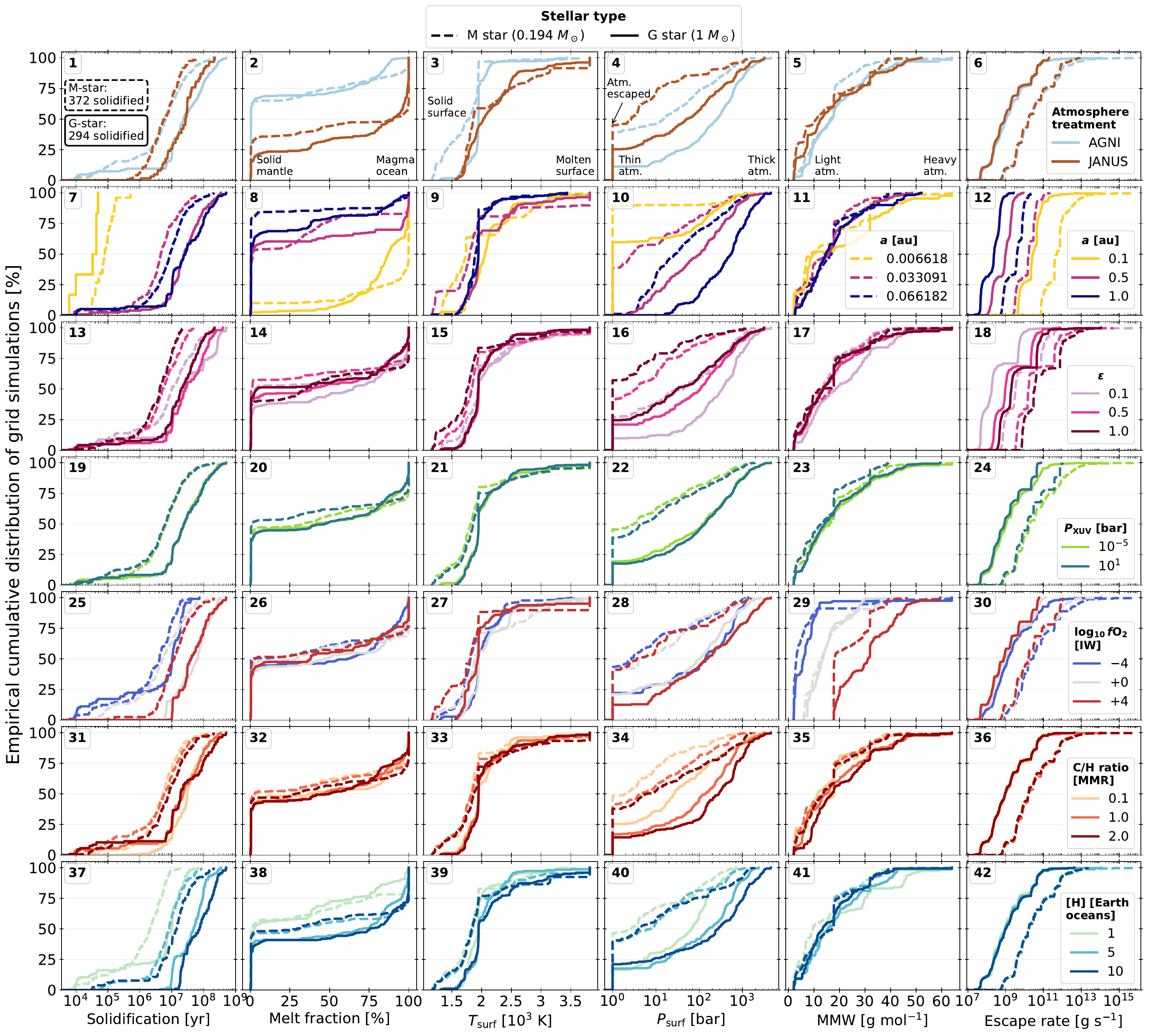}
    \caption{Empirical cumulative distribution function of solidification time, melt fraction, surface temperature, surface pressure, mean molecular weight and escape rate for eight input parameters explored in the simulation grid for planets orbiting a $1\,M_\oplus$ star (solid lines) and a $0.194\,M_\oplus$ star (dashed lines), including atmospheric escape. Each row represents a tested value of an input parameter, indicated by its color in the legend, while each column shows the corresponding output of the simulation at the final time step. The first column, showing magma ocean solidification time, displays only simulations that end up with a solidified status (294 cases around a G star, 372 around an M star) while all other panels represents all outcomes of our grid (radiative equilibrium, volatiles escaped, mantle solidified). All subplots share the same y-axis, representing an empirical cumulative distribution in percent.}
    \label{ecdf grid 2 star}
\end{figure*}

Figure\,\ref{ecdf grid 2 star} presents results of our simulation grid for a Sun-like star (solid lines) and an M dwarf (dashed lines) using normalized empirical cumulative distributions. We varied eight input parameters (Table\,\ref{table grid parameters}): stellar mass,  atmosphere model, semimajor axis $a$, XUV reference pressure $P_\mathrm{XUV}$, escape efficiency factor $\epsilon$, upper-mantle oxygen fugacity $\Delta$IW, bulk C/H ratio, and bulk hydrogen inventory. The figure empirically shows how several output variables respond to variations in these input variables of our grid. The examined output quantities are measured at the end of each simulation. The shape and relative locations of the distribution curves in each panel together quantify the sensitivity of each dependent variable (columns) to each independent variable (rows). 

In the first column, we plot the magma ocean crystallization time for 294 (G star) and 372 (M star) of the simulated planets that ended due to complete mantle solidification. A radiative-convective treatment (\texttt{AGNI}, light blue lines in panel 1) shortens crystallization times by several orders of magnitude for a subset of our planets (starting at $10^4$\,years) whereas in a purely convective atmosphere (\texttt{JANUS}, brown lines), even the fastest-solidifying planets take at least 1\,Myr.
Orbital separation---varied in the second row of the figure---is a second dominant factor on magma ocean crystallization time (first column), regardless of the stellar type. For short separations ($a$=0.1\,au and $a$=0.006618\,au; gold lines in panel 7), only a small fraction of planets fully solidify: in our sample, 6 out of 258 planets around Sun-like stars and 26/271 around M dwarfs fully crystallize. The majority of close-orbit planets remain in a long-lived magma ocean state, with about 90\% of the simulation with a melt fraction higher than 20\% for close-in planets (panel 8, gold lines). Close-in planets are subject to intense stellar irradiation, with equilibrium temperatures of $T_{\rm eq} = 910$\,K (around G star) and $T_{\rm eq} = 1100$\,K (around M dwarf). When coupled with high mean molecular weight atmospheres, the greenhouse effect can help sustain long-lived magma oceans on those planets. More distant bodies ($a$=0.5\,au and $a$=0.03309\,au, $a$=1.0\,au and $a$=0.066182\,au; pink and dark blue lines in panel 7, respectively) exhibit longer cooling periods for more than 90\% of the sample: from 1\,Myr up to 0.5\,Gyr. Planets on close-in orbits experience stronger atmospheric escape (gold lines, panel 12), which rapidly removes most of their atmospheric content. The subsequent absence of a dense atmosphere allows the planets to crystallize more quickly. For Earth‑like orbits ($a$=1.0\,au, dark blue solid line in panel 7), crystallization times shorten from a few Myr down to 0.01\,Myr when considering a radiative-convective atmosphere (brown solid line in panel 1) for 10\% of simulated planets. 
Variations in escape efficiency $\epsilon$ (panel 13) influence crystallization times, whereas changes in  $P_\mathrm{XUV}$ (panel 19)---which defines the radius $R_\mathrm{XUV}$ where hydrodynamic escape begins---have a negligible impact on crystallization times. In particular, a low escape efficiency ($\epsilon=0.1$, light pink lines, panel 13) prolongs atmospheric survival and thereby delays mantle crystallization, especially for planets orbiting M dwarfs. Planets with a large hydrogen inventory (10\,Earth oceans, dark blue lines in panel 37) solidify on longer timescales (from 100\,Myr up to 0.5\,Gyr) than those with a lower hydrogen content (from $10^4$\,years up to $4 \times 10^7$\,years). Hydrogen-rich planets exhibit a larger volatile reservoir, resulting in a thicker atmosphere and higher surface pressures (panel 40), which together reduce radiative heat loss and prolong the magma ocean phase \citep[as per][]{elkins2012magma, Hamano-2015,Nicholls_2024_JGRP}. 

While the first column considers only planets that have fully solidified, the remaining columns include all planets that have either reached radiative equilibrium, solidified, or lost their volatile inventories through atmospheric escape. 
Melt fraction is most sensitive to atmospheric treatment and orbital separation. Planets with a purely convective atmosphere (\texttt{JANUS}; brown lines in panel 2) end with higher melt fractions than those with a radiative-convective atmosphere. 20\% of simulations around a G star with a purely convective atmosphere solidified (30\% around an M star), while it is almost 70\% when considering radiative layers in the atmosphere. Not accounting for radiative layers leads to less efficient atmospheric cooling, thus prolonging the magma ocean phase. Melt fraction increases with decreasing semimajor axis (panel 8): about 70\% of planets at 1.0\,au are fully solid by the end of the simulation, whereas only a few percent of those at 0.1\,au reach a solid state. More than 90\% of close-in planets maintain melt fractions $\Phi > 50\%$, meaning that over half of their mantles remain molten at termination time. These planets receive intense stellar irradiation that, when combined with atmospheric blanketing effects, can maintain high surface temperatures and inhibit mantle cooling. Across the entire simulation grid, atmospheric energy-transport treatment and orbital separation emerge as the dominant controls on magma ocean evolution.

Surface temperatures range from about $1150$\,K to nearly $3800$\,K across all simulated planets (third column panels). About 65\% of the simulations around a G star (and 80\% around an M dwarf) exhibit $T_\mathrm{surf} < 2000$\,K corresponding to cold surfaces of partially molten or fully solidified planets. 35\% of G star simulations (and 20\% for M dwarf) end up with a global magma ocean: $T_\mathrm{surf} > 2000$\,K and $\Phi > 0.95$). 

Surface pressure primarily depends on orbital separation (panel 10) but also on the initial volatile content of the planet: volatile-poor planets develop lower surface pressures (carbon-poor, beige lines in panel 34; hydrogen-poor, light green lines in panel 40). 60\% of the simulations at 0.1\,au (gold solid line, panel 10) have a 1\,bar surface pressure ($P_{\rm surf}=P_{\rm stop} = 1\,\mathrm{bar}$) due to intense XUV radiations leading to atmosphere depletion at the last time step, consistent with efficient escape ($\epsilon=1$, dark purple lines in panel 16). Around a Sun-like star, 25\% of \texttt{JANUS} simulations terminate due to volatile depletion, compared to only 10\% for \texttt{AGNI} cases. Radiative layers in the atmosphere accelerate cooling of both the atmosphere and the planet’s interior, leading to magma ocean solidification before significant atmospheric loss.

Previous studies have demonstrated that mantle redox state governs volatile outgassing, playing a primary role in setting atmospheric composition \citep{Burgisser_2015, Gaillard_2015, Gaillard_2022, Deng_2020, Ortenzi_2020, Liggins_2022, Tian_2024, BRACHMANN2025, Gkouvelis_2025}. Our simulations show that this relationship is preserved when atmospheric escape is included. As illustrated in panel\,29, the final atmospheric mean molecular weight (MMW) reflects the initial mantle redox state, consistent with previous studies \citep{Burgisser_2015, Gaillard_2022, Deng_2020, Ortenzi_2020, Gkouvelis_2025}. Reduced initial conditions ($f$O$_2$\,=\,IW-4, blue lines) produce lighter atmospheres, with 95\% of simulations ending with MMW\,$< 16\,\mathrm{g\,mol^{-1}}$. Oxidized states ($f$O$_2$\,=\,IW+4, red lines) yield to much heavier atmospheres spanning from $18\,\mathrm{g\,mol^{-1}}$ up to $64\,\mathrm{g\,mol^{-1}}$, with only the heaviest atmospheres being sulfur-dominated. 

Escape rates are controlled primarily by semimajor axis (panel 12) and escape efficiency (panel 18). Planets at 0.006618\,au (gold dashed line; panel 12) experience the strongest mass-loss rates---up to $10^{16}\,\mathrm{g\,s^{-1}}$---because they receive intense stellar XUV irradiation, whereas for planets at 1\,au, the highest mass-loss rates are about $\sim\,10^{9}\,\mathrm{g\,s^{-1}}$. This results in a difference of up to seven orders of magnitude in escape rates between close-in and distant planets. The wide range of escape efficiencies, $\epsilon = 0.1$--$1$, significantly amplifies this spread, with the most efficient cases ($\epsilon = 1$, dark purple lines; panel 18) producing extreme escape rates exceeding $10^{12}\,\mathrm{g\,s^{-1}}$ for distant planets, up to $10^{16}\,\mathrm{g\,s^{-1}}$ for close-in planets. For context, present‑day atmospheric escape rates on Earth \citep[$\sim\,3\times 10^{3}\,\mathrm{g\,s^{-1}}$;][]{Catling_Kasting_2017, Gronoff_2020} and Venus \citep[$\sim\,10^{3}\,\mathrm{g\,s^{-1}}$;][]{Lammer_2006, Gronoff_2020} are dominated by non‑hydrodynamic processes such as Jeans escape, charge exchange, polar, ion/sputtering escape, photochemical escape, and unmagnetized ion outflow. Such escape rates are many orders of magnitude lower than the energy‑limited rates in our simulations. The step-like patterns seen across the escape-rate panels arise from the discrete sets of $a$ and $\epsilon$ values (and other tested parameters) sampled in the simulation grid.

\subsubsection{Atmosphere retention as a function of escape rates}

\begin{figure}
    \resizebox{\hsize}{!}{\includegraphics{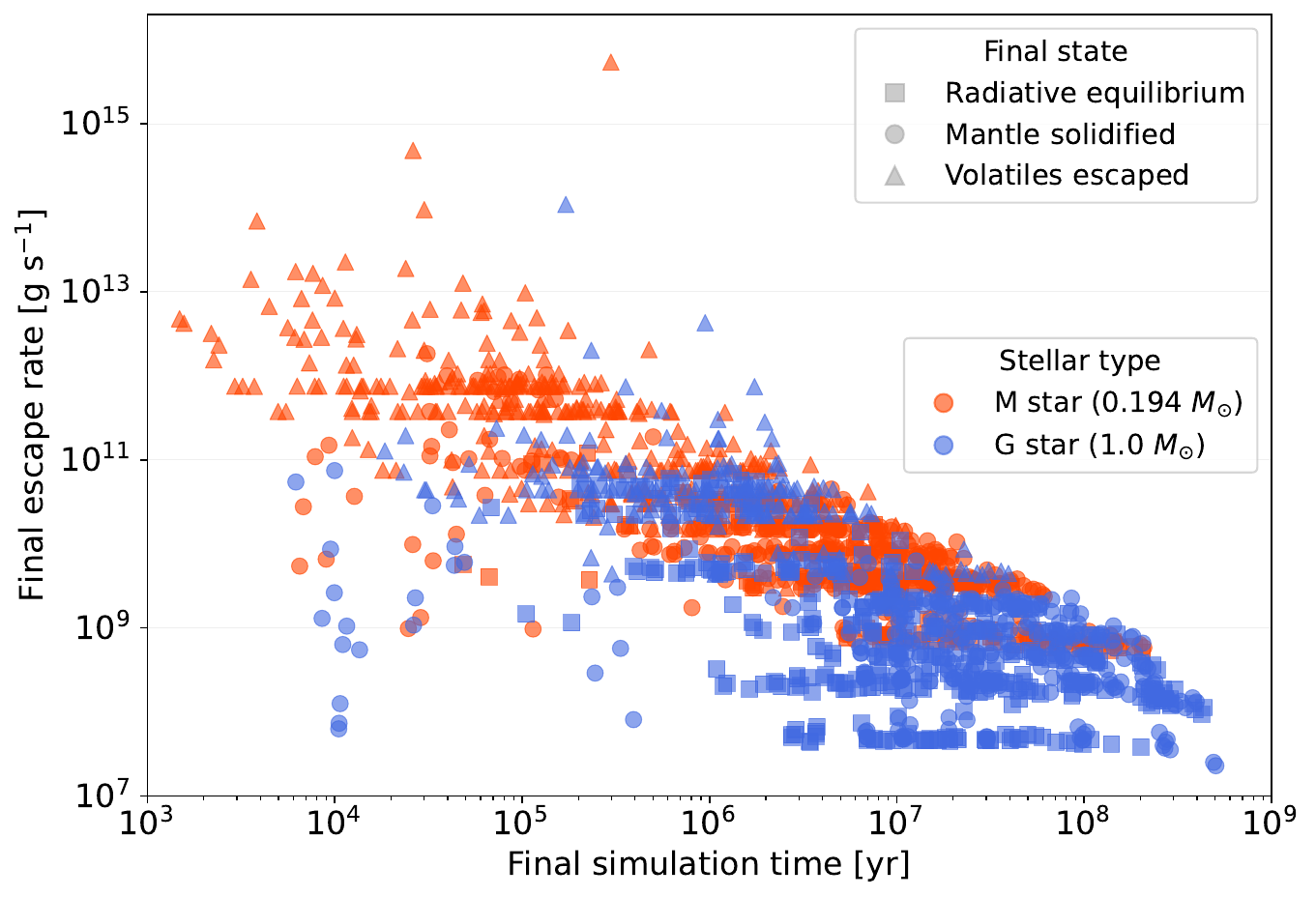}}
    \caption{Final escape mass loss rate at the end of the simulation for planets orbiting an M star (red points) or a G star (blue points). Simulations terminate due to atmospheric escape (triangles), mantle solidification (circles), or radiative equilibrium (squares).}
    \label{plot escape rates}
\end{figure}

We examine how atmospheric escape rates affect the survival of planetary atmospheres in Fig.\,\ref{plot escape rates}, highlighting the contrasting retention outcomes for planets orbiting M-dwarf and G-type stars. Planets experiencing extreme escape rates ($> 10^{12}\,\mathrm{g\,s^{-1}}$) lose their entire atmosphere (triangles) on very short timescales: within $10^{6}$\,years. Those cases are predominantly found in M-dwarf systems (upper-left cluster of red points).
Planets that survive until mantle solidification (circles) or reach radiative equilibrium (squares) occupy the lower-right region of the plot. These planets exhibit significantly lower escape rates ($< 10^{10}\,\mathrm{g\,s^{-1}}$) and evolve over much longer timescales, up to $\approx\,0.5\,$\,Gyr. Their ability to retain an atmosphere at the end of the simulation highlights two distinct regimes: one dominated by atmospheric retention at low escape rates and another dominated by atmospheric loss at high escape rates. Planets with escape rates above $\sim\,10^{12}\,\mathrm{g\,s^{-1}}$ experience rapid atmospheric loss and short evolutionary timescales, particularly around M dwarfs, whereas planets with escape rates below $\sim\,10^{10}\,\mathrm{g\,s^{-1}}$ can retain their atmospheres and evolve until mantle solidification or radiative equilibrium.

\subsubsection{Controls on atmospheric mean molecular weight}
\label{MMW section results}

\begin{figure}
    \resizebox{\hsize}{!}{\includegraphics{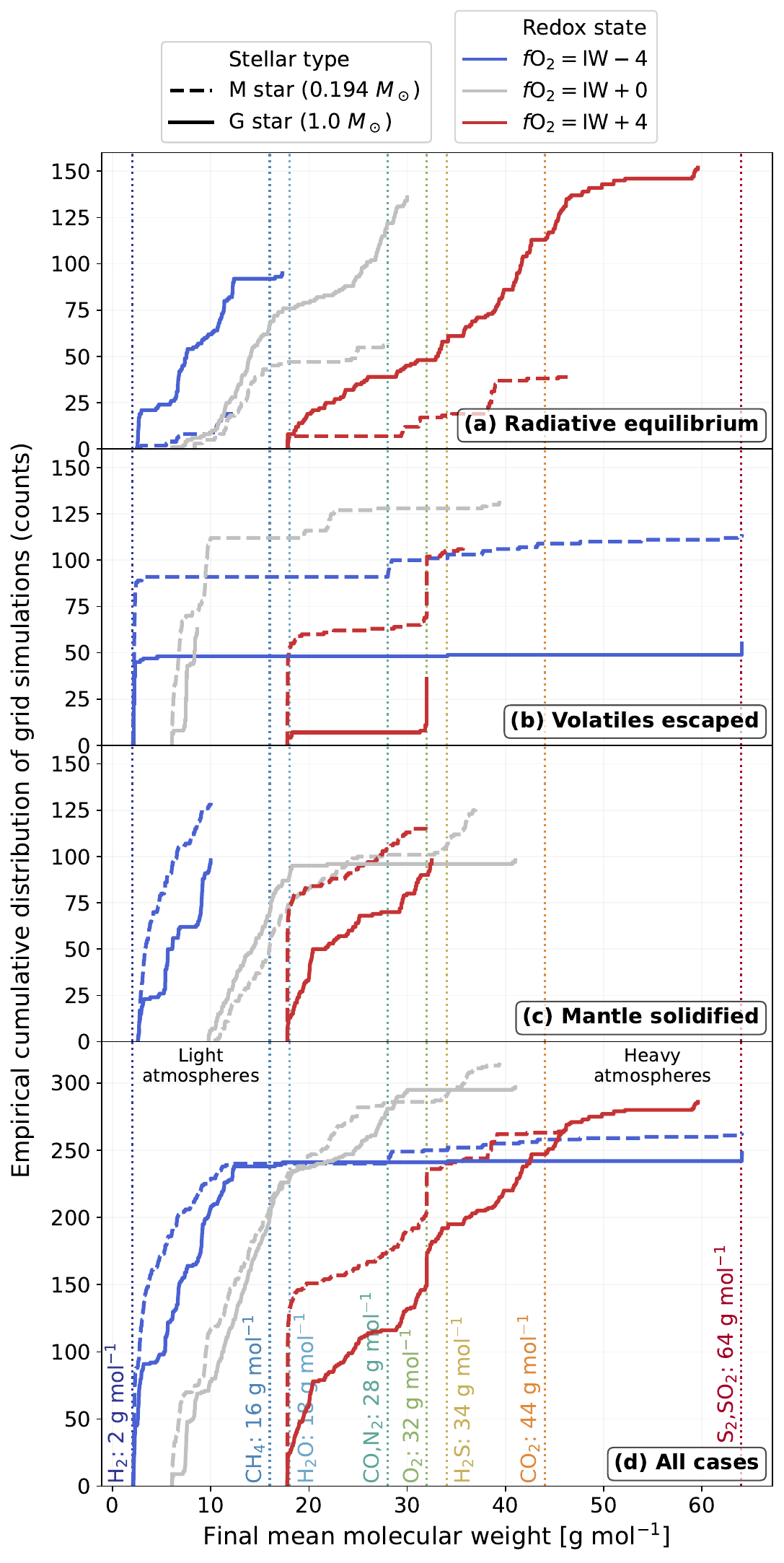}}
    \caption{Empirical cumulative distribution function of final mean molecular weight across the escape grid for three oxygen fugacities ($\Delta$IW=[-4,\,+0,\,+4]), separated by termination criterion: radiative equilibrium reached (a), complete volatile loss (b), mantle solidification (c), and all cases combined (d). Solid lines corresponds to simulations with a Sun-like star and dashed lines for an M star. Vertical dotted lines indicate the molecular weight of peculiar molecules (\element{H_2}, \element{CH_4}, \element{H_2O}, \element{CO}, \element{N_2}, \element{O_2}, \element{H_2S}, \element{CO_2}, \element{S_2} and \element{SO_2}) present in the final atmospheres of our simulations. Panel (d) is comparable to panel 29 in Fig.\,\ref{ecdf grid 2 star}.}
    \label{distribution MMW}
\end{figure}

\begin{figure*}
    \centering
    \includegraphics[width=\textwidth]{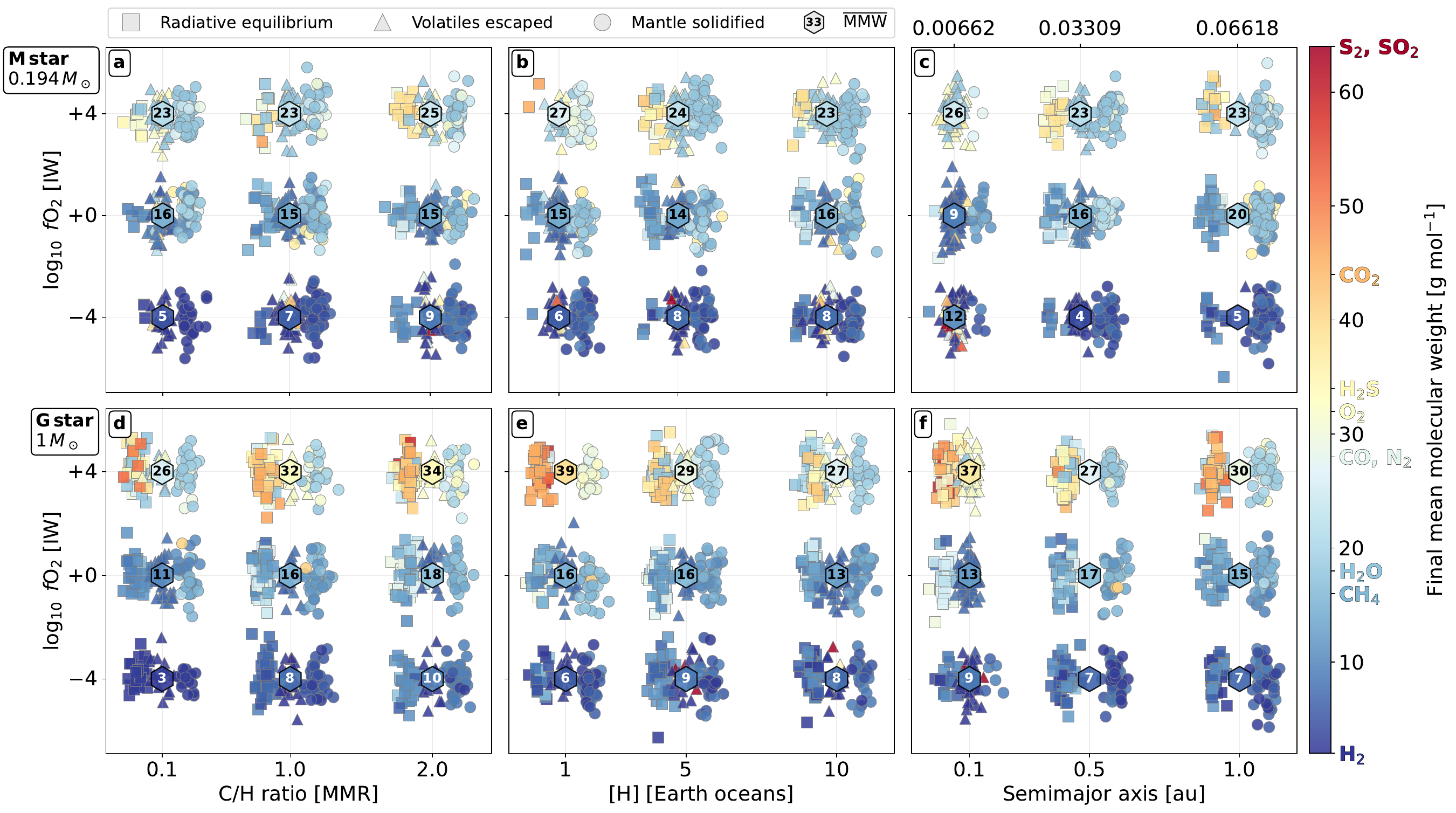}
    \caption{Final atmospheric mean molecular weight as a function of upper-mantle oxygen fugacity (y-axis), volatile budgets (C/H ratio and hydrogen reservoir), and semimajor axis for two stellar types: M-dwarf (top panels) and G-type (bottom panels). Squares indicate simulations that reached radiative equilibrium; triangles mark cases where atmospheric volatiles were fully lost; circles denote scenarios in which the mantle solidified. To improve visibility, points within each cluster are slightly offset both vertically and horizontally. All final simulated atmosphere is colored by its final MMW (color bar). Hexagonal markers show the mean molecular weight averaged ($\overline{\rm MMW}$) over each cluster of points corresponding to a given oxygen fugacity and parameter set, with the numerical value displayed at the center of each hexagon.}
    \label{6 panels exploration mmw}
\end{figure*}

Figure\,\ref{distribution MMW} shows the cumulative distributions of the final mean molecular weight for three different oxygen fugacities and two stellar types across our escape grid, based on model termination. Potential atmospheric compositions are indicated by dotted lines, with colors corresponding to their molecular weights.

Stellar type is the primary driver of the planet's evolutionary outcome. Planets orbiting G-type stars (solid lines) are significantly more likely to reach radiative equilibrium (383 cases) compared to those orbiting M dwarfs (117 cases), especially for planets with an oxidized mantle (panel a; red lines). An oxidized mantle replenishes the atmosphere with heavier volatiles such as \element{CO_2} and \element{H_2O} which produce a strong greenhouse effect that traps heat near the surface and maintains a permanent magma ocean. For a planet to reach a steady state while retaining an atmosphere, atmospheric replenishment needs to be sufficiently efficient to offset atmospheric escape. 
The higher XUV luminosities of M dwarfs drives higher escape rates (Fig.\,\ref{plot escape rates}) and more frequent model termination due to atmospheric loss compared to Sun-like systems (Fig.\,\ref{distribution MMW}, panel b). As a result, M-dwarf planets are 2\,times more likely to terminate due to total volatile depletion than planets around Sun-like stars. 
Additionally, more planets solidify around M-dwarf hosts (372 planets) than G-type stars (294 planets) because rapid atmospheric escape removes heat, accelerating mantle cooling (Fig.\,\ref{distribution MMW}, panel c). 

While stellar type shapes the planet’s evolutionary pathway, the final atmospheric composition remains strongly influenced by the initial mantle redox state, consistent with previous studies \citep[panel d,][]{Gaillard_2015, Deng_2020, Ortenzi_2020, BRACHMANN2025, Gkouvelis_2025}. 
Notably, while the $f$O$_2$-MMW relationship is well established, very few planets with a reduced mantle (blue lines) undergoing atmospheric depletion (panel d) develop transiently sulfur-rich atmospheres in their final stages---a composition that would not emerge from outgassing models alone. Such planets reach a MMW up to $64\,\mathrm{g\,mol^{-1}}$ just before complete atmospheric loss, regardless of stellar type. Volatile species such as sulfur are highly soluble in a reduced mantle \citep{Namur_2016,Dasgupta_2022}, making them among the last gases to be outgassed before atmospheric removal. This suggests that the final stage of atmospheric escape for planets with reduced mantles may be enriched in sulfur-bearing species. This result highlights the evolutionary transition of dominant atmospheric species, showing how atmospheres can evolve from light, \element{H_2}-rich compositions to heavier, sulfur-rich states, reflecting the role of reduced mantles in shaping exoplanet atmospheres as volatile loss progresses.
For oxidized planets that solidify (panel c, red lines), water outgassing dominates, with a significant fraction of simulations (70\%) producing atmospheres $<\,18\,\mathrm{g\,mol^{-1}}$ for planets orbiting M-dwarf hosts. Oxidized planets orbiting M dwarfs that reach radiative equilibrium (red dashed line; panel a) retain atmospheres with mean molecular weights as high as $\approx48\,\mathrm{g\,mol^{-1}}$, while planets around Sun-like stars exhibit even heavier atmospheres, with MMW reaching $\approx60\,\mathrm{g\,mol^{-1}}$ (noting that these end states correspond to different simulation timescales; see Fig.\,\ref{plot escape rates}). This difference arises because the stronger XUV irradiation around M dwarfs drives more efficient atmospheric escape, causing a larger fraction of planets to lose most or all of their atmospheres (Fig.\,\ref{distribution MMW}, panel b) on relatively short timescales (Fig.\,\ref{plot escape rates}, red triangles).
Carbon-bearing species are generally less soluble in silicate melts than hydrogen-bearing species and therefore outgas earlier during magma-ocean evolution \citep{Bower_2022}. 
However, atmospheric escape removes atmospheres more efficiently from planets orbiting M dwarfs, resulting in a larger fraction of planets around G-type stars retaining heavier atmospheres for radiative-equilibrium cases.

Figure\,\ref{6 panels exploration mmw} shows how final atmospheric mean molecular weight varies across our simulated planets grid for two stellar mass, as a function of initial volatile budget: C/H ratio (panels a and d), hydrogen content (panel b and e), and semimajor axis (panel c and e) for three mantle redox states (IW–4, IW+0, IW+4). Each symbol indicate model termination status of a single simulation, colored by its end‐state atmospheric MMW (color bar; blue = light species, red = heavy species). 

As expected from previous studies \citep{Gaillard_2015, Deng_2020,Ortenzi_2020,Gkouvelis_2025}, initial oxygen fugacity remains the most critical factor that determines final atmospheric mean molecular weight---a relationship that is preserved even when atmospheric escape is included---regardless of stellar type. Oxidized planets orbiting Sun-like stars exhibits the highest mean molecular weight ($44\,\mathrm{to}\,64\,\mathrm{g\,mol^{-1}}$) for low hydrogen content (1\,Earth ocean of hydrogen, panel e) when the planets reach radiative equilibrium (red squares). Those planets indicate atmospheres dominated by heavy, oxidized species like \element{CO_2} or \element{SO_2}. 
Planets with a reduced mantle result in low mean molecular weight atmospheres: $3\,\mathrm{to}\,12\,\mathrm{g\,mol^{-1}}$, dominated by light species as \element{H_2} regardless of the host star. 

The initial volatile budget and orbital separation function as secondary controls. The initial volatile inventory can shift the final MMW by about $10\,\mathrm{g\,mol^{-1}}$, resulting in corresponding variations in the atmospheric scale height and surface pressure. For a fixed redox state ($f$O$_2$\,=\,IW--4), varying the C/H content strongly impacts the final MMW: it increases the average of mean molecular weight ($\overline{\rm MMW}$) from $3\,\mathrm{g\,mol^{-1}}$ for C/H\,=\,0.1 up to $\overline{\rm MMW} = 10\,\mathrm{g\,mol^{-1}}$ for a carbon-rich initial abundance (lower line in panel d). A similar behavior is observed when varying the initial hydrogen inventory for an oxidized mantle (upper line in panel e): low hydrogen content leads to $\overline{\rm MMW} = 39\,\mathrm{g\,mol^{-1}}$, whereas an initial reservoir of 10\,Earth oceans reduces it to $\overline{\rm MMW} = 27\,\mathrm{g\,mol^{-1}}$.
Low hydrogen inventory (equivalent to 1\,Earth ocean of hydrogen; panel e) and carbon-rich initial conditions (C/H\,=\,2.0; panel d) lead to the heaviest atmospheres when starting with an oxidized mantle for Sun-like star systems ($\overline{\rm MMW} = 39\,\mathrm{g\,mol^{-1}}$ and $\overline{\rm MMW} = 34\,\mathrm{g\,mol^{-1}}$ respectively). In contrast, low mean molecular weight atmospheres, appear for low initial carbon content (C/H\,=\,0.1; panel d) and reduced mantles ($\overline{\rm MMW} = 3\,\mathrm{g\,mol^{-1}}$; dark blue markers in panel d). 
Semimajor axis strongly influences atmospheric escape and therefore model termination. In G-star systems (panel f), only the innermost planets at 0.1\,au lose their atmospheres, independent of mantle redox state. A similar pattern appears for M-dwarf hosts: planets on very close-in (0.00662\,au) and intermediate (0.03309\,au) orbits experience efficient atmospheric removal. Reduced, close-in planets around M dwarfs (panel c, lower-left cluster) almost always undergo complete atmospheric loss. As suggested by Fig.\,\ref{distribution MMW}b, the final stage of atmospheric evolution prior to full escape appears to be sulfur-rich, as indicated by the yellow (\element{H_2S}), orange (\element{CO_2}) and red (\element{SO_2}, \element{S_2}) triangles marking the last surviving atmospheres in panel c.
For an oxidized planet orbiting a Sun-like star at 1.0\,au (panel f, top-right corner), planets end up in two different regimes: either planets solidify and retain a water-dominated atmosphere, or they reach radiative equilibrium and exhibits a \element{CO_2}-dominated atmosphere on top of a permanent magma ocean. 

\subsubsection{Influence of escape processes on final atmospheric composition}

\begin{figure}
    \resizebox{\hsize}{!}{\includegraphics[width=\hsize]{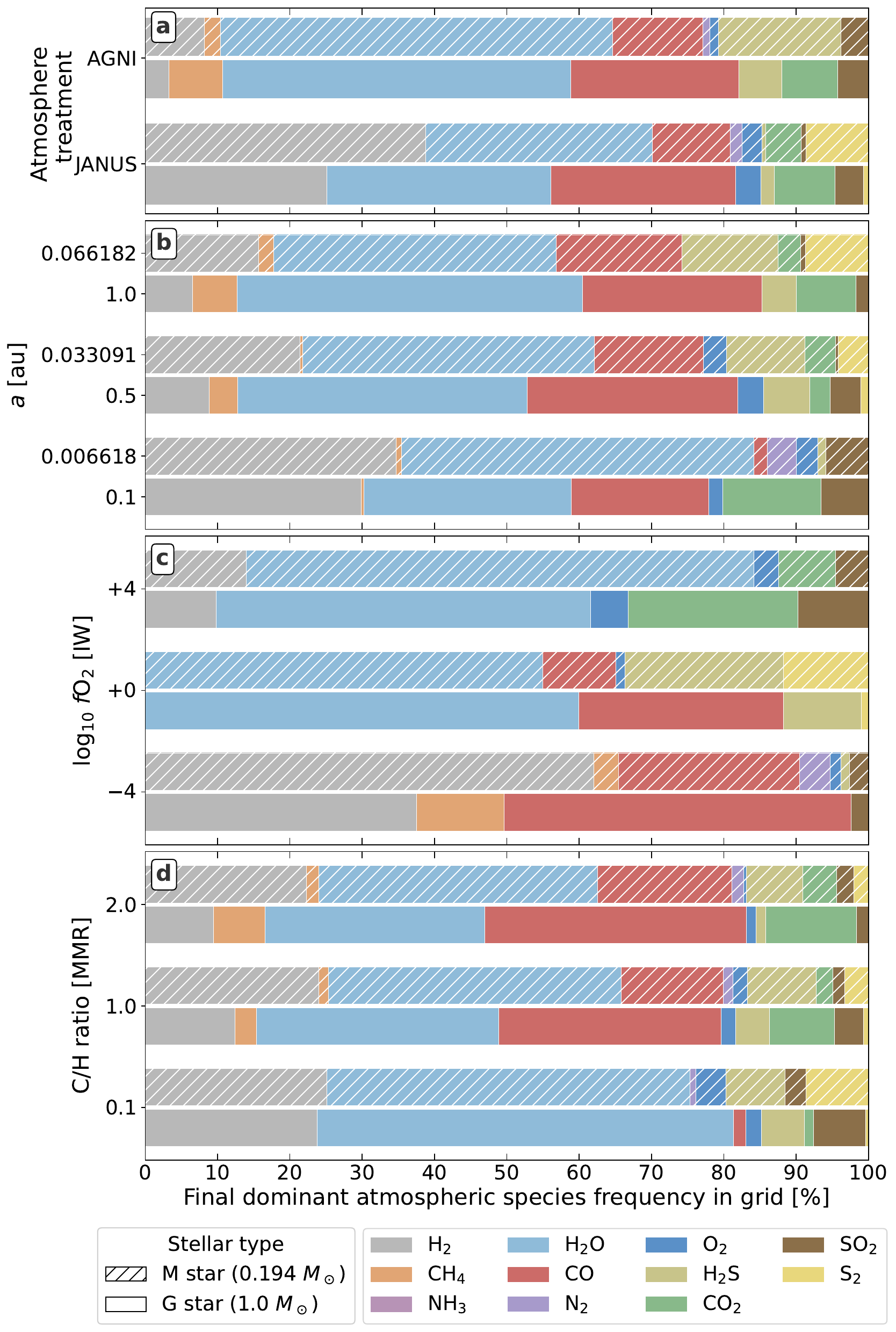}}
    \caption{Final dominant atmospheric species by mass across simulation grid as a function of atmosphere treatment (a), semimajor axis (b), oxygen fugacity (c), and C/H ratio (d). For each input parameter tested in our grid, bars indicate the percentage of simulations sharing that parameter value in which the most abundant species dominates the atmosphere, aggregated over all other initial conditions. We included all converged simulations, comprising cases where the interior solidified, reached radiative equilibrium, or where the atmosphere has escaped; in the last case, the reported dominant species corresponds to the last time step. Results are shown separately for planets orbiting M-dwarf stars (hatch style) and Sun-like stars.}
    \label{final atm composition}
\end{figure}

Figure\,\ref{final atm composition} presents the final dominant species by mass in the atmosphere (regardless of model termination) as a percentage across all converged grid simulations. Across all grid simulations, \element{H_2} (gray), \element{H_2O} (light blue), and \element{CO} (red) emerge as the most common final atmospheric species, strongly dependent on initial mantle redox state, with water consistently being the most abundant at the end of the simulations.

Including radiative layers in the atmosphere (or not) significantly influences the final atmospheric composition. When the atmosphere is assumed to be fully convective (\texttt{JANUS}, panel a), a larger fraction of simulations ends with \element{H_2}-dominated atmospheres (39\% for M dwarfs; panel a). In contrast, simulations that allow radiative layers (\texttt{AGNI}) produce fewer \element{H_2}-dominated outcomes (8\% for M dwarfs), favoring water-rich atmospheres instead (54\%). The development of radiative layers accelerates atmospheric cooling near the surface, which impacts magma ocean evolution. Consequently, most \texttt{AGNI} simulations terminate due to magma ocean crystallization, releasing water stored in the molten mantle and resulting in predominantly \element{H_2O}-dominated atmospheres. As temperatures decline, thermochemical equilibrium shifts carbon speciation from \element{CO} toward \element{CO_2} or \element{CH_4} (depending on redox conditions) and sulfur speciation from \element{S_2} to \element{H_2S}, consistent with the relative fractions obtained in the simulations (e.g., \element{CO_2} and \element{CH_4} each accounting for 8\% around G stars and \element{H_2S} accounting for 17\% of M-dwarf outcomes with \texttt{AGNI}).

Orbital separation also has a significant influence on the final atmospheric composition (panel b). Planets in close orbits ($a$=0.1\,au around G stars, $a$=0.0066\,au around M dwarfs) tend to have lighter, \element{H_2}-rich atmospheres (30--35\%). Sun-like stars exhibit heavier atmospheres, including \element{CO} (19\%), \element{CO_2} (14\%), or \element{SO_2} (7\%) as major species, whereas 49\% of M dwarf simulations end with water-dominated atmospheres due to efficient water outgassing during magma ocean crystallization. At larger orbital separations, water and carbon monoxide remain significant, but heavier atmospheres are more commonly \element{H_2S}- and \element{S_2}-dominated. Methane-dominated atmospheres are primarily found at large orbital separations, accounting for roughly 6\% of the simulations, under reducing conditions and following planetary cooling.

Mantle redox state remains a key parameter in shaping final atmospheric composition even when including atmospheric escape (panel c). In a reduced environment, 62\% of M dwarf simulations and 38\% of Sun-like star simulations end with hydrogen-dominated atmospheres. \element{CO} is the other dominant species, appearing in 25\% of M dwarf cases and 48\% of Sun-like stars. Methane is present only in reduced environments, more frequently around Sun-like stars (12\%) than M dwarfs (3\%). A small fraction (3\%) of simulations end \element{SO_2}-dominated, corresponding to late-stage sulfur-driven escape. For $f$O$_2$\,=\,IW+0, more than half of simulations produce \element{H_2O}-dominated atmospheres, regardless of stellar type, due to efficient water outgassing. \element{H_2S} is more prevalent around M dwarfs (22\%) than G stars (10\%), while \element{S_2} dominates 12\% of M dwarf cases. \element{CO}-dominated atmospheres are more common around G stars (28\%). Oxidized mantle ($f$O$_2$\,=\,IW+4) exhibit mostly water-rich atmospheres (70\% for M dwarfs, 52\% for Sun-like stars). For Sun-like stars, \element{CO_2} (24\%) and \element{SO_2} (10\%) also appear as dominant species. \element{H_2}-dominated atmospheres are rare (10--15\%), likely because most \element{H_2O} remains dissolved in a long-lived magma ocean, while \element{H_2}, which is poorly soluble, appears in the atmosphere even during prolonged magma ocean stages. This accounts for the greater fraction of \element{H_2}-rich atmospheres in \texttt{JANUS} (39\% for M star) compared to \texttt{AGNI} simulations (8\%, panel a).

A high C/H ratio causes the apparition of C bearing atmospheres (panel d). For \element{CO}-dominated atmospheres, C/H ratios above 1.0 lead to \element{CO} appearing in up to 36\% of Sun-like star simulations and 19\% of M dwarf simulations. Methane and \element{CO_2} also require a high initial carbon content ($\mathrm{C/H}>1.0$).

Overall, mantle redox state is a key control on the final atmospheric composition across our grid, and this dependence is preserved even when atmospheric escape is included. The resulting distributions are shaped by outgassing governed by solubility laws and by hydrodynamic escape. 
Three main regimes arose from our simulations: \element{H_2}-, \element{H_2O}- and \element{CO}-dominated atmosphere representing respectively 24\%, 43\% and 12\% of M dwarf systems and 15\%, 40\% and 24\% of G star systems. These patterns arise from magma ocean solubility: \element{H_2}, poorly soluble, accumulates in the atmosphere even when the mantle is molten, whereas \element{H_2O} and \element{CO} remain largely retained in the melt until crystallization. Heavier atmospheres are rare ($<10\%$) and form under conditions that favor sulfur or carbon accumulation: \element{S_2} and \element{H_2S} dominate at large orbital separations with moderately reduced mantles ($f$O$_2$\,=\,IW+0), while \element{CO_2} and, to a lesser extent \element{SO_2}, appear in close-in planets with oxidized mantles.

\section{Discussion}
\label{discussion section}

Currently JWST spectroscopic capabilities allow us to probe the atmospheres of dominantly gaseous and low-density exoplanets \citep{Benneke_2024,Dyrek_2024, Hu_2024,Piaulet-Ghorayeb_2024,teske_2025}. A key open question is whether small rocky planets, particularly those orbiting M-dwarf hosts, can retain atmospheres over long timescales \citep{Owen_Mohanty_2016, Wunderlich_2019,Krissansen-Totton-2024,kreidberg_stevenson_2025}. Our simulations provide theoretical context for how volatile rich atmospheres can be built and eroded during early planetary evolution around G-type and M-dwarf stars.

\subsection{Validity of the energy-limited approximation}

\subsubsection{Uncertainties in XUV fluxes}
\label{section fxuv uncertainties}

Modeling stellar XUV emission remains challenging because of ISM absorption and uncertainties in stellar rotational evolution \citep{Richey-Yowell-2019-hazmat, Peacock_2020, Tu-2015, Johnstone_2021}. In particular, the duration of the saturated XUV phase depends strongly on stellar mass and initial rotation, ranging from $\approx10$--$300$\,Myr for G stars \citep{Tu-2015} to up to $\approx1$\,Gyr for fully convective M dwarfs \citep{kippenhahn1990stellar, Johnstone_2021}. Since hydrodynamic escape is most efficient during this phase, different stellar evolution models can produce integrated XUV fluxes that differ by factors of $\approx2$--$10$ \citep{Kubyshkina-2018, Johnstone_2021}, leading to substantially different atmospheric mass-loss histories.

Thus, our calculated mass loss rates represent an upper bound. In reality, lower XUV fluxes would result in reduced escape rates, implying that atmospheric removal at short orbital separations may not be systematic, potentially allowing for alternative evolutionary pathways where atmospheres are retained. \citet{Kubyshkina_Vidotto_2022} demonstrated that similar planets can end up as either bare cores or sub-Neptunes depending purely on the star's rotation history.

\subsubsection{Constraints on escape efficiency}
\label{discussion epsilon section}

We explored a broad parameter space spanning from $\epsilon=0.1$ up to 1.0 in Fig.\,\ref{ecdf grid 2 star} (panels 13-18), making it important to assess whether those values remain consistent with physical conditions. Reported $\epsilon$ values differ substantially across studies, reflecting their strong dependence on planetary mass, which controls gravitational escape and atmospheric composition, which affects radiative efficiency. Hot Jupiters with pure hydrogen atmospheres usually undergo very efficient hydrodynamic escape, with $\epsilon \approx 0.6$ \citep[for a Jovian atmosphere]{Waite-1983,Yelle_2004}, whereas the hydrogen-rich atmosphere on the early Earth is often modeled with $\epsilon = 0.3$ \citep{Watson_1981}. \citet{Wordsworth_2018} investigated escape from low-mass planets using $0.15 < \epsilon < 0.3$ and \citet{Kasting-1983} adopted $\epsilon = 0.15$ to model escape from a water-rich early Venus. Several studies have extended this range to investigate highly efficient escape scenarios: \citet{Luger_Barnes_2015} considered values up to 0.30 for terrestrial planets around M dwarfs and \citet{Owen_Wu_2013, Koskinen-2014, Lehmer_Catling_2017} explored efficiencies up to 0.60 to simulate intense escape. However, \citet{Shematovich-2014} showed that for hydrogen-dominated atmospheres in gas giants, values of $\epsilon$ above 0.20 tend to overestimate hydrogen escape rates. \citet{Owen-2012} argued the same for super-Earths, i.e., $\epsilon$ should not exceed 0.20. For super-Earths and rocky planets, $\epsilon = 0.15$ is consistent with a number of studies \citep{Kasting-1983, Shematovich-2014, Luger_Barnes_2015, Erkaev-2016, Kubyshkina-2018, Owen_2019, Cherubim_2024}.

Our results for $\epsilon > 0.3$ likely represent extreme escape scenarios for atmospheric retention. Our models indicate that atmospheres are less likely to survive under high-efficiency conditions (dark purple lines in panel 18, Fig.\,\ref{ecdf grid 2 star}) as predicted by \citet{Owen_Wu_2017}. Therefore, atmospheric survival is more realistic under lower efficiency values ($\epsilon\approx\,0.15$). The upper bound of our range ($\epsilon=1.0$) represents the adiabatic limit where all received $F_\mathrm{XUV}$ is perfectly converted into the kinetic energy required for the atmosphere to escape. But other processes like radiative cooling, where the heated atmospheric gas re-emits energy back into space, act as energy sinks, significantly reducing the total XUV energy that can be converted into energy-limited efficiency \citep{Owen_Wu_2013, Shematovich-2014, Nakayama-2022,Chatterjee_2024,Yoshida-2024}. \citet{Yoshida-2024} showed that radiative cooling by carbon species can suppress hydrodynamic escape in \element{H_2}-rich atmospheres on Earth-sized planets, while \citet{Nakayama-2022} demonstrated more generally that thermochemical and radiative cooling processes can significantly reduce the heating efficiency and mass-loss rates. Together, these studies indicate that such processes can extend atmospheric lifetimes. Because our escape model did not account for radiative cooling, $\epsilon = 1.0$ represents a non-physical upper limit on the mass-loss rate and should be viewed as an extreme, largely unrealistic regime.

Furthermore, \citet{Owen-2012} showed that treating $\epsilon$ as a constant in time is a simplification. The efficiency is dynamic, evolving with the mass, radius of the planet over time, and the changing incident flux. For super-Earths, $\epsilon$ typically declines during atmospheric evolution, as ongoing mass loss removes atmospheric mass and limits the conversion of XUV energy into escape. This implies that our fixed-$\epsilon$ models might overestimate mass loss in the late stages of evolution, as atmospheric composition evolves more gradually and magma ocean solidification proceeds over long timescales, reducing escape efficiency compared to earlier stages. Despite this, adopting $\epsilon < 0.3$ as a baseline for rocky planets provides a conservative estimate that aligns with the majority of atmospheric escape simulations \citep{Kasting-1983, Luger_Barnes_2015}; although $\epsilon$ controls the instantaneous mass loss rate, it has a negligible impact on the final MMW, surface temperature or mantle melt fraction in our models (see Fig.\,\ref{ecdf grid 2 star}).

\subsubsection{Non-thermal escape and atmospheric fractionation}

\begin{figure}
    \resizebox{\hsize}{!}{\includegraphics{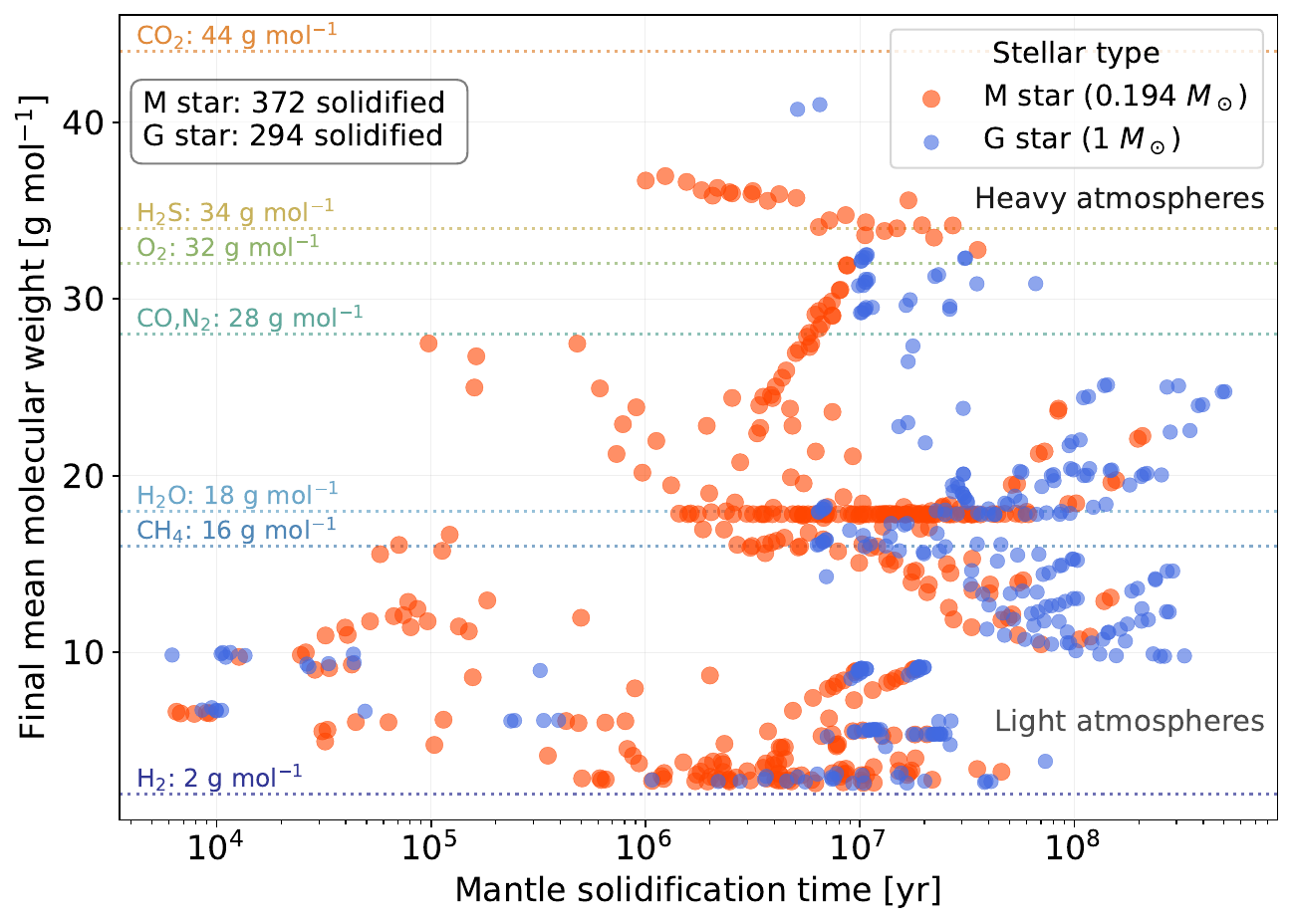}}
    \caption{Final mean molecular weight of rocky planets atmospheres as a function of mantle solidification time for planets orbiting M-dwarf (orange) and G-type (blue) stars. Horizontal dotted lines indicate the molecular weight of common volatiles (\element{H_2}, \element{CH_4}, \element{H_2O}, \element{CO}, \element{N_2}, \element{O_2}, \element{H_2S}, \element{CO_2}).}
    \label{plot discussion MMW MO solid}
\end{figure}

In the high-irradiation regime characteristic of the early stellar lifetimes considered here, hydrodynamic outflow is the dominant loss mechanism \citep{Watson_1981, Lammer-2003}. Kinetic processes, such as Jeans escape and non-thermal loss mechanisms, are negligible compared to the massive bulk outflow driven by XUV radiation and are therefore not included in this model. \citet{Kislyakova_2014} showed that the maximum non-thermal escape rate of pick-up ionized hydrogen atoms for the super-Earth Kepler\,11\,b ($1.97\,R_\oplus$, $4.3\,M_\oplus$) is $\approx7\times10^7\,\mathrm{g\,s^{-1}}$. This value is comparable to the lower bound of the thermal (hydrodynamic) escape rates obtained in our simulations for $1\,M_\oplus$ planets, namely $\approx2\times10^7\,\mathrm{g\,s^{-1}}$ around a young Sun-like star and $\approx5\times10^8\,\mathrm{g\,s^{-1}}$ around a young M dwarf (Fig.\,\ref{plot escape rates}). However, only $\approx11\%$ of the Sun-like star cases yield hydrodynamic escape rates $<10^8\,\mathrm{g\,s^{-1}}$, within the same range as the non-thermal escape estimated by \citet{Kislyakova_2014}. In most of our simulations, hydrodynamic escape dominates, supporting the neglect of non-thermal processes.

Atmospheric escape can fractionate hydrodynamic outflows under strong XUV irradiation, which photodissociates molecules such as \element{H_2O} and allows hydrogen to escape more efficiently than heavier species \citep{Zahnle_Kasting_1986, Wordsworth_2018, Cherubim_2024, Cherubim_2025}. Preferential loss of light atoms enriches the atmosphere in heavier constituents over time, leading to secondary atmospheres dominated by species such as \element{H_2O}, \element{CO}, \element{CO_2}, and \element{O_2} \citep{Johnstone_2020, Cherubim_2025}. The particle fluxes in our simulations generally exceeded the critical flux required to drag heavy species, following \citet{Wordsworth_2018, Cherubim_2024}. We adopted the prescription of \citet{Yoshida_2022}, using a threshold value for \element{H_2O} in a \element{H_2} background of $1.9\times10^8\,\mathrm{g\,s^{-1}}$. Most of our simulated atmospheres became water-dominated, with 43\% of planets around M stars and 40\% around Sun-like stars exceeding this threshold. In the Sun-like star simulations, 19\% of planets fell below this threshold and entered a fractionated escape regime, whereas all M-star planets remained above it. Overall, our simulations remained largely within the valid regime of energy-limited bulk escape.

In our approach, only volatiles that have been outgassed into the atmosphere are removed by escape, effectively treating the atmosphere as a fractionated volatile reservoir for the planet. In \texttt{PROTEUS}, escape occurs uniformly across all atmospheric species at a fixed total mass-loss rate. For example, if $\dot{M}_\mathrm{EL} = 10\,\mathrm{g\,s^{-1}}$ for an atmosphere composed of equal mass mixing ratios of \element{H} and \element{N}, then $5\,\mathrm{g\,s^{-1}}$ of hydrogen and $5\,\mathrm{g\,s^{-1}}$ of nitrogen are removed during that time step, where elemental mass mixing ratios are derived from molecular molar mixing ratios.

Our bulk-escape treatment provides a conservative prescription for the evolution of atmospheric mean molecular weight (MMW). Figure\,\ref{plot discussion MMW MO solid} shows that atmospheres with higher MMW tend to be associated with longer mantle solidification times. Including fractionation would preferentially remove hydrogen and leave behind oxygen-enriched atmospheres \citep{Cherubim_2024}, yielding to higher final MMW than predicted by our bulk-escape approach. Accounting for fractionation would therefore further delay magma ocean solidification, particularly for planets orbiting M dwarfs. Our neglect of fractionation thus implies that both the predicted magma ocean lifetimes and final atmospheric MMW represent conservative lower bounds. This behavior is illustrated in Fig.\,\ref{final atm composition}, where \element{H_2}, which is poorly soluble in the mantle, accumulates early in the atmosphere in \texttt{JANUS} simulations, while \element{H_2O} and \element{CO}, retained in the melt until crystallization, dominate the final atmospheric compositions in most \texttt{AGNI} cases.

\subsection{Dispersal of the atmosphere}

Our evolutionary models demonstrated that atmospheric escape is a critical factor, reducing the longevity of atmospheres on rocky planets, with the most significant effects observed around M-dwarf stars, as shown by the higher escape rates in Fig.\,\ref{plot escape rates} \citep{Luger_Barnes_2015, Johnstone-2019}. In these systems, extreme cases show atmosphere stripping in only $10^4$\,years, whereas planets around G-type stars may retain their atmospheres up to 0.5\,Gyr in our simulations. Nevertheless, atmosphere dispersal within only $10^4$\,years is far shorter than usual rapid escape timescale: $10^5$--$10^6$\,years \citep{Owen_Mohanty_2016,Johnstone-2019} and is therefore highly unlikely. Such rapid atmospheric loss arises only in the cases where the planet is placed in an extreme hydrodynamic-escape regime (i.e $\epsilon = 1.0$), which in our simulations corresponds to the most extreme and physically limiting end of the explored parameter space.
Planets orbiting G-stars experience lower XUV radiation than M-dwarfs over their evolution \citep{Ribas_2005, Tu-2015}, resulting in less efficient atmospheric escape rates. Our simulations, shown in Fig.\,\ref{ecdf grid 2 star} (panel 10), confirmed that for the same instellation $\mathcal{S}$, planets orbiting M-dwarfs lose their atmospheres at orbital separations corresponding to $0.1\,\mathcal{S}_\oplus$ (where $\mathcal{S}_\oplus$ denotes today’s Earth instellation) and $0.5\,\mathcal{S}_\oplus$ (i.e., $a$=0.00662\,au and $a$=0.03309\,au in our grid) while Sun-like systems only lost their atmospheres for the close-in cases ($a$=0.1\,au). Consistent with earlier work \citep{Owen-2012, Kubyshkina-2018, Johnstone-2019}, our results similarly indicate that M-dwarf planets, especially those near the inner edge of the habitable zone, are generally unable to retain substantial secondary atmospheres. Consistent with \citet{Hamano_2013}, we found that planets orbiting farther out ($a$=1.0\,au) around a Sun-like star experience conditions that favor atmospheric retention during the modeled evolutionary phase, particularly if radiative equilibrium or mantle solidification is achieved (Fig.\,\ref{6 panels exploration mmw}, panel f). However, atmospheric loss at later times remains possible beyond the scope of these simulations. We found that atmospheres can be retained for moderate escape rates ($< 10^{10}\,\mathrm{g\,s^{-1}}$, Fig.\,\ref{plot escape rates}) where the escape flux is balanced by continuous volatile replenishment from the mantle for $1\,M_\oplus$ planets. Additionally, purely convective atmospheres can extend the magma ocean lifetime up to a permanent state (Fig.\,\ref{TP_profiles_escape_ON_AGNI_JANUS}), consistent with previous study by \citet{Nicholls_2025_MNRAS}. This arises from a reduced radiative cooling efficiency compared to radiative-convective models, which may also increase the tendency for atmospheric retention, although purely convective atmospheres are less physically realistic than treatments like \texttt{AGNI}.

A variety of atmospheric compositions emerged in our simulations, as shown by Fig.\,\ref{final atm composition}, ranging from hydrogen‐rich atmospheres to \element{H_2O}- or \element{CO}-mixtures and even to Venus‐like, \element{CO_2}-dominated atmospheres. Such a diversity in atmosphere compositions is consistent with previous magma ocean outgassing studies \citep{ElkinsTanton2008, Herbort2020, Solomatova2021}. In the cases where the magma ocean solidifies (Fig\,\ref{plot discussion MMW MO solid}), the remaining atmosphere tends toward a high mean molecular weight, in agreement with magma ocean degassing models \citep{ElkinsTanton2008, Solomatova2021}. When planets reach radiative equilibrium (Fig\,\ref{distribution MMW}, panel a), heavier volatile species---such as \element{H_2S}, \element{CO}, \element{CO_2}---are preferentially outgassed for oxidized mantles, increasing the mean molecular weight of the residual atmosphere from $\approx\,30\,\mathrm{g\,mol^{-1}}$ up to $60\,\mathrm{g\,mol^{-1}}$, consistent with redox-dependent outgassing trends reported by \citet{Gaillard_2015}, \citet{Deng_2020}, \citet{Ortenzi_2020}, \citet{BRACHMANN2025} and \citet{Gkouvelis_2025}. Reduced conditions retained lower mean molecular weight atmospheres (\element{H_2}- up to \element{H_2O}-dominated atmospheres), as also found by \citet{Herbort2020} and \citet{Gaillard_2022}.

\subsection{Longevity of magma ocean phases}

In our simulations, magma‐ocean lifetimes span from roughly 0.01\,Myr to 0.5\,Gyr for the subset of planets that reach full crystallization (35\% of Sun-like systems and  44\% of M-dwarf systems in our models; see Fig.\,\ref{plot discussion MMW MO solid}). Predicted magma ocean lifetimes span a similarly broad range, $10^4$--$10^7$\,years, reflecting different assumptions about atmospheric opacity, volatile budgets, redox buffering and interior-atmosphere coupling \citep{Hamano_2013,Krissansen_Totton_2021, Nicholls_2024_JGRP,seidler2024,Cherubim_2025}. Our findings are consistent with those of \citet{Nicholls_2024_JGRP}, which showed that increasing planetary hydrogen content prolongs the mantle‐crystallization timescale up to $\approx\,10\,$\,Myr (Fig.\,\ref{ecdf grid 2 star}, panel 37). 

Atmospheric structure exerts a strong control on cooling efficiency: the presence of a radiative (non-convective) layer modifies the outgoing long wave radiation and can either accelerate or retard solidification depending on the atmosphere’s opacity and composition \citep[Fig.\,\ref{TP_profiles_escape_ON_AGNI_JANUS},][]{Lebrun2013, Marcq2017, Lichtenberg_2021_JGRP, 2023selsis,cmiel_Character_2025, Nicholls_2025_MNRAS}. For planets around Sun-like stars, 232 with atmospheric radiative layers solidify compared to 62 with purely convective atmospheres; for M-dwarfs, the numbers are 251 versus 121. 

We found that orbital separation exerts a strong control on solidification time for close-in planets, supported by \citet{Hamano_2013} and \citet{Nicholls_2024_JGRP}. Only 6 planets at 0.1\,au solidify, compared to 104 and 121 at 0.5\,au and 1.0\,au respectively, with a similar trend for M-dwarf systems. Short-lived magma oceans occur on close-in planets with low initial hydrogen inventories and reduced mantles, particularly when atmospheric radiative layers enhance cooling. Conversely, the longest-lived---and in some cases permanent---magma oceans arise on planets at larger orbital separations with oxidized mantles, high hydrogen inventories, moderate escape rates, and convective atmospheres that evolve toward radiative equilibrium.

\subsection{Storage of volatiles in the planetary interior}

Under highly oxidizing conditions ($f$O$_2$\,=\,IW+4), an Earth‐sized planet at 0.1\,au around a G‐type star retains primarily \element{H_2O} and \element{S_2} in its magma ocean throughout most of its evolution (Fig.\,\ref{evolution escape on off}, panel f). Upon complete solidification, \element{CO_2} is outgassed in modest amounts, increasing its mole fraction to $\approx\,10\%$ at the surface. The retention of \element{H_2O} during the magma‐ocean phase, due to water’s high solubility in silicate melts \citep{Mysen2014}, aligns with previous studies \citep{Cherubim_2025} with \element{H_2O} being the dominant species in the planetary interior of oxidized mantle. Sulfur species (\element{S_2} and \element{SO_2}) constitute the second-largest volatile reservoir available for atmospheric replenishment, followed by carbon-bearing species at lower abundance.

Our model assumed homogeneous mantle crystallization proceeding from the core--mantle boundary upward to the surface. Under this assumption, all trapped volatiles are released at final solidification. Therefore, our results set an upper limit on the volatile content in the resulting atmosphere. In reality, a heterogeneous crystallization scenario is more likely for rocky planets---forming isolated melt pockets that trap volatiles in the deep mantle over geological timescales \citep{Hier-Majumder_Hirschmann_2017,Bower_2018, Bower_2022,Lichtenberg_2025}. A more complex, non-monotonic solidification history could trap a fraction of the heavy, oxidized species in the mantle \citep{Moore_2023,Lichtenberg_2025}. Such scenario would produce a less oxidized atmosphere with a lower mean molecular weight. It would also allow for later-stage outgassing as crystallization and mantle overturn proceed \citep{Kite-2020,Guimond_2024,Baumeister_2025}.

In \texttt{SPIDER}, we computed the mantle liquidus and solidus (phase boundaries describing the transition from a solid to a liquid) only for \element{MgSiO_3}, being the principal mantle component. However, \citet{Cherubim_2025} showed that high‐solubility volatiles such as water can remain dissolved inside the melt during several Myr. If the interior volatile reservoir grows significantly over the planet's evolution, it would alter silicate melting relations and shift both the liquidus and solidus curves toward lower temperatures. Future work needs to incorporate \element{H_2O} phase in melting curves to capture these feedbacks over evolutionary timescales for atmosphere-interior interactions.

The dissolution of volatiles in the magma ocean partially counterbalances atmospheric escape by regulating exchange between the atmosphere and the interior. Highly soluble species such as \element{H_2O} and \element{S_2} can remain dissolved in the mantle or be outgassed and subsequently ingassed back into the magma ocean \citep{Dorn_lichtenberg_2021,Salvador_2023_fluid, Carone_2025}. In this way, ingassing temporarily reduces atmospheric volatile abundances by transferring species into the melt, thereby slowing net atmospheric loss. These volatiles can later be reintroduced to the atmosphere through continued outgassing, modifying atmospheric composition during the late stages of planetary evolution and sustaining atmospheres through late-stage magma ocean activity \citep{Sossi_2020, Dorn_lichtenberg_2021, Cherubim_2025}. While atmospheric escape remains a dominant driver of atmospheric evolution, our simulations showed that exchange with the magma ocean fractionates the planet’s volatile budget and extends the timescale over which an atmosphere can be maintained.

\subsection{Model assumptions}
\label{discussion model assumptions}

We adopted radiative equilibrium as one of the termination criteria for our simulations. While atmospheric escape and interior processes may continue to modify planetary properties beyond this point, we focused on the early stages of planetary evolution, where the majority of volatile outgassing and escape occurs. Terminating the simulations at radiative equilibrium captures the main processes shaping early atmospheric composition and volatile budgets. This approach avoids additional uncertainties related to long-term evolution, including solid-state convection, surface condensation, and cloud feedback. These processes shall be treated in future works.

We did not include thermochemical processes in the lower atmosphere, above the surface, or photochemistry in the upper atmosphere in our simulations. However, these chemical processes will be addressed in future work. Incorporating photochemistry could also  account  for the formation of hazes and aerosols, altering the opacity of atmospheric layers and thus the thermal structure of the upper atmosphere \citep[Fig.\,\ref{TP_profiles_escape_ON_AGNI_JANUS};][]{Hu_2012, Tsai_2021, kitzmann_2023}. Those improvements would help determine which species are present at the XUV‑absorption level ($P_\mathrm{XUV}$) and their abundances, enabling a more realistic treatment of atmospheric escape. In future work, coupling our escape model to a full photochemical network and accounting for fractionation will be essential to capture this feedback. These processes may subtly modify escape rates: photochemistry could enhance escape by generating lighter species at high altitudes, whereas haze formation may slightly reduce escape by increasing upper‑atmosphere opacity.

By neglecting clouds and hazes, our model assumed a fully clear-sky atmosphere, allowing direct stellar radiation to be absorbed or reflected by Rayleigh scattering. Condensation of volatiles is allowed, but in the absence of clouds or hazes, the atmospheric opacity is dominated by gaseous species. However, previous studies showed that cloud or haze layers can significantly alter both the $P$--$T$ profile and radiative transfer, by changing the penetration depth of stellar flux \citep{Marley_2013, Lavvas_2021}. On the other hand, thick cloud decks can introduce an additional greenhouse effect that would offset the scattering albedo contribution of the cloud. As a result, our work should be interpreted as applying to an idealized clear‑sky scenario, while atmospheres with clouds/hazes may exhibit different evolutionary pathways depending on the composition and radiative properties of the aerosols formed.

\citet{Nicholls_2025_tidal_l9869} demonstrate that including tidal heating in evolution models of close-in rocky worlds around M‐dwarf (for the L\,98‑59 system) prolongs magma‐ocean crystallization by two orders of magnitude. When combined with atmospheric escape, tidal heating could shorten magma‐ocean lifetimes for close‐in planets, potentially altering both their thermal history and volatile inventories. In our simulations, all planets have zero eccentricity ($e$ = 0), which implies that stellar-driven tidal dissipation is negligible. Tidal heating requires varying tidal stresses (due to orbital eccentricity or obliquity) to dissipate energy into heat \citep{Jackson_2008, Henning_2009, Driscoll_2015,Farhat2025ApJ}. As a result, neglecting tidal heating is a realistic approximation for our modeled planets across a wide orbital range (0.1--1\,au for a Sun-like star and 0.00662--0.06618\,au for an M dwarf) and is unlikely to affect our main conclusions.

\section{Conclusions}
\label{conclusions}

We presented a comprehensive parameter study of the coupled interior-atmosphere evolution of molten rocky planets, focusing on the role of hydrodynamic atmospheric escape during these early magma ocean phases. Using the \texttt{PROTEUS} framework, we explored a wide parameter space of 1674 converged simulations for Earth-sized planets around Sun-like stars (1\,$M_\odot$) and M dwarfs (0.194\,$M_\odot$), incorporating atmospheric loss via the \texttt{ZEPHYRUS} module. 

Our key findings are:

\begin{enumerate}
       
    \item Efficient atmospheric escape and a radiative-convective atmospheric structure together shorten magma ocean lifetimes by weakening greenhouse insulation and enhancing heat loss (Fig.\,\ref{evolution escape on off} and Fig.\,\ref{TP_profiles_escape_ON_AGNI_JANUS}). Mantle solidification may be accelerated by up to several orders of magnitude, depending on a planet's orbital separation and its volatile distribution (Fig.\,\ref{ecdf grid 2 star}).
      
    \item Atmospheric escape plays a key role in the evolution of magma ocean planets, as it can remove the atmosphere and significantly shorten magma ocean lifetimes (Fig.\,\ref{plot discussion MMW MO solid}). For 1\,$M_{\oplus}$ planets orbiting at warm and temperate separations, atmospheres can survive if escape rates remain moderate, below $10^{10}\,\mathrm{g\,s^{-1}}$ during planetary evolution (Fig.\,\ref{plot escape rates}). Otherwise, the atmosphere is lost and rapid magma ocean solidification exposes a bare rocky surface.
      
    \item When atmospheric escape is included, mantle redox state still controls atmospheric composition via outgassing, with interior oxygen fugacity  influencing the final mean molecular weight (Fig.\,\ref{6 panels exploration mmw}) and dominant species (Fig.\,\ref{final atm composition}), consistent with previous studies \citep{Deng_2020, Ortenzi_2020, Gaillard_2022, Gkouvelis_2025}.
      
    \item Diverse evolutionary pathways emerge from variations in stellar type, initial volatile inventory, orbital separation, mantle redox state, and atmospheric escape efficiency, resulting in a wide range of end states---from bare rocky planets to thick atmospheres above magma oceans (Fig.\,\ref{distribution MMW})---with compositions ranging from \element{H_2}- to \element{SO_2}-dominated (Fig.\,\ref{final atm composition}).
\end{enumerate}

Our simulations of rocky planets around both G-type (1\,$M_\odot$) and M-dwarf (0.194\,$M_\odot$) stars offer a theoretical perspective on the ongoing debate over whether planets, particularly those orbiting active M-dwarfs, can retain atmospheres. We emphasize on the importance of exchange with the magma ocean and long-term outgassing in sustaining their atmospheres over long timescales. 
Future work will include implementing escape fractionation into our coupled-interior atmosphere framework, to allow direct connections between escape processes and the observed atmospheric compositions of rocky (exo)planets. Those additions will improve interpretations of observable atmospheres and better inform upcoming spectroscopic missions targeting rocky exoplanets such as the Extremely Large Telescope \citep[ELT;][]{Kasper-2021}, the Large Interferometer for Exoplanets \citep[LIFE;][]{Quanz-2022}, and the Habitable Worlds Observatory \citep[HWO;][]{Harada-2024,Mamajek-2024}.

\section*{Code and data availability}

All computer codes used in this work are open source software available on GitHub:
\begin{itemize}
    \item \texttt{PROTEUS: }\url{https://github.com/FormingWorlds/PROTEUS}
    \item \texttt{SPIDER: }\url{https://github.com/djbower/spider}
    \item \texttt{CALLIOPE: }\url{https://github.com/FormingWorlds/CALLIOPE}
    \item \texttt{JANUS: }\url{https://github.com/FormingWorlds/JANUS}
    \item \texttt{AGNI: }\url{https://github.com/nichollsh/AGNI}
    \item \texttt{SOCRATES: }\url{https://github.com/nichollsh/SOCRATES}
    \item \texttt{MORS: }\url{https://github.com/FormingWorlds/MORS}
    \item \texttt{ZEPHYRUS: }\url{https://github.com/FormingWorlds/ZEPHYRUS}
\end{itemize}
Data and plots produced for this work are available on Zenodo at \url{https://doi.org/10.5281/zenodo.18485113}.
\begin{acknowledgements}

We thank our anonymous reviewers for their positive and constructive feedback which significantly improved this work. The authors thank Mara Attia and Tadahiro Kimura for valuable discussions on atmospheric escape. This work was supported by the Branco Weiss Foundation, the Netherlands eScience Center (PROTEUS project, NLESC.OEC.2023.017), the Alfred P. Sloan Foundation (AEThER project, G-2025-25284), NASA’s Nexus for Exoplanet System Science research coordination network (Alien Earths project, 80NSSC21K0593), and the NWO NWA-ORC PRELIFE Consortium (PRELIFE project, NWA.1630.23.013). H.N. acknowledges support from STFC grant UKRI1184. The simulations presented in this work made use of the H\'abr\'ok high performance computing cluster, enabled by the Center for Information Technology of the University of Groningen. The authors made use of OpenAI’s ChatGPT (GPT-5 mini) for code optimization and language refinement. 

\end{acknowledgements}

\bibliographystyle{aa} 
\bibliography{bibliography_doc}
\end{document}